\documentclass[twocolumn,astrosym]{aastex62}

\graphicspath{{./}{}}

\received{2018 September 11}
\revised{2018 December 17}
\accepted{2019 January 15}
\published{2019 February 28}
\submitjournal{AJ}

\shorttitle{RGZ: Radio Galaxy Distortion by Clusters}
\shortauthors{Garon et al.}

\usepackage{amsmath}

\begin{document}

\title{Radio Galaxy Zoo: The Distortion of Radio Galaxies by Galaxy Clusters}

\correspondingauthor{Avery F. Garon}
\email{garo0040@umn.edu}

\author[0000-0001-9409-0892]{Avery F. Garon}
\affil{Minnesota Institute for Astrophysics, School of Physics and Astronomy, University of
Minnesota, 116 Church St. SE, \\Minneapolis, MN 55455, USA}

\author[0000-0001-5636-7213]{Lawrence Rudnick}
\affil{Minnesota Institute for Astrophysics, School of Physics and Astronomy, University of
Minnesota, 116 Church St. SE, \\Minneapolis, MN 55455, USA}

\author[0000-0002-2504-7628]{O. Ivy Wong}
\affil{International Centre for Radio Astronomy Research-M468, The University of Western Australia,
35 Stirling Hwy, Crawley, \\WA 6009, Australia}

\author[0000-0002-9368-4418]{Tom W. Jones}
\affil{Minnesota Institute for Astrophysics, School of Physics and Astronomy, University of
Minnesota, 116 Church St. SE, \\Minneapolis, MN 55455, USA}

\author[0000-0001-8380-9988]{Jin-Ah Kim}
\affil{Minnesota Institute for Astrophysics, School of Physics and Astronomy, University of
Minnesota, 116 Church St. SE, \\Minneapolis, MN 55455, USA}
\affil{Department of Astronomy \& Center for Galaxy Evolution Research, Yonsei University, Seoul
03722, Republic of Korea}

\author[0000-0003-4873-1681]{Heinz Andernach}
\affil{Departamento de Astronom\'ia, DCNE, Universidad de Guanajuato, Apdo. Postal 144, CP 36000,
Guanajuato, Gto., Mexico}

\author[0000-0001-5064-0493]{Stanislav S. Shabala}
\affil{School of Natural Sciences, University of Tasmania, Private Bag 37, Hobart, Tasmania 7001,
Australia}

\author[0000-0002-5289-5729]{Anna D. Kapi{\'n}ska}
\affil{National Radio Astronomy Observatory, 1003 Lopezville Rd, Socorro, NM 87801, USA}

\author[0000-0002-4597-1906]{Ray P. Norris}
\affil{CSIRO Astronomy and Space Science, Marsfield, NSW 2122, Australia}
\affil{Western Sydney University, Locked Bag 1797, Penrith, NSW 2751, Australia}

\author[0000-0003-4439-2627]{Francesco de Gasperin}
\affil{Hamburger Sternwarte, Universit\"at Hamburg, Gojenbergsweg 112, D-21029, Hamburg, Germany}

\author[0000-0002-5787-1726]{Jean Tate}
\affil{Zooniverse Citizen Scientist}
\affil{Oxford Astrophysics, Denys Wilkinson Building, Keble Road, Oxford OX1 3RH, UK}

\author[0000-0002-7300-9239]{Hongming Tang}
\affil{School of Physics and Astronomy, University of Manchester, Oxford Road, Manchester, M13 9PL,
UK}



\begin{abstract}

We study the impact of cluster environment on the morphology of a sample of 4304 extended radio
galaxies from
Radio Galaxy Zoo. A total of 87\% of the sample lies within a projected 15~Mpc of an optically
identified cluster. Brightest cluster galaxies (BCGs) are more likely than other cluster members to
be radio sources, and are also moderately bent. The surface density as a function of separation from
cluster center of non-BCG radio galaxies follows a power law with index $−1.10\pm 0.03$ out to
$10~r_{500}$ ($\sim 7$~Mpc), which is steeper than the corresponding distribution for optically
selected galaxies. Non-BCG radio galaxies are statistically more bent the closer they are to the
cluster center. Within the inner $1.5~r_{500}$ ($\sim 1$~Mpc) of a cluster, non-BCG radio galaxies
are statistically more bent in high-mass clusters than in low-mass clusters. Together, we find that
non-BCG sources are statistically more bent in environments that exert greater ram pressure. We use
the orientation of bent radio galaxies as an indicator of galaxy orbits and find that they are
preferentially in radial orbits. Away from clusters, there is a large population of bent radio
galaxies, limiting their use as cluster locators; however, they are still located within
statistically overdense regions. We investigate the asymmetry in the tail length of sources that
have their tails aligned along the
radius vector from the cluster center, and find that the length of the inward-pointing tail is
weakly suppressed for sources close to the center of the cluster.

\end{abstract}



\section{Introduction}

Radio galaxies exhibit a wide range of morphological variation, especially those that are members of
galaxy clusters. As radio-emitting jets from the galactic nucleus expand into the surrounding
environment, they interact with the diffuse plasma of the intracluster medium (ICM). Density
variations within the ICM and relative motion between the radio galaxy and the ICM, for example, can
produce distortions in the radio emissions and more complicated structures.
 
From simulations, we have some understanding of the dynamical processes that underlie the
interaction between the ICM and the radio jets
\citep[e.g.][]{balsara92,loken95,heinz06,porter09,mendygral12}. In individual galaxy clusters, we
can use observations of radio galaxies to map shocks, turbulence, and other structures within the
ICM \citep[e.g.][]{krempec-krygier95,roettiger99,feretti12,owen14,degasperin17}. However, the more
general dependence of radio galaxy structures on the physical characteristics of the ICM remain an
open question, one which case studies cannot answer.
 
To this end, we require a very large sample of radio galaxies. By distributing the classification
work across a large number of citizen scientists, the Radio Galaxy Zoo
project\footnote{\url{radio.galaxyzoo.org}} \citep[RGZ;][]{banfield15} is generating a catalog of
almost 180,000 radio source host identifications. Using an initial portion consisting of 118,090
radio sources, we investigate how we can use the morphology and orientation of the radio emission to
probe the influence of the ICM as a function of its inferred properties.

In this paper, we address this question from three main fronts. The first is by investigating how
the radio galaxies are bent by the ICM. As galaxies travel through the ICM, ram pressure will
distort the radio jets \citep{begelman79,jones79,jones17}. Ram pressure depends on the density of
the surrounding medium and the relative velocity between the galaxy and the ICM, so the amount of
bending we observe will be a function of these quantities. The orientation of the angle bisector of
the opening angle depends on the direction of relative motion between the galaxy and the ICM, so the
second line of investigation is based on mapping the orbits of radio galaxies by using the bending
as a tracer of galaxy orbits. Finally, the two radio jets of a single galaxy expand into different
parts of the ICM, so by observing asymmetry in the jets we will characterize density and pressure
gradients in the ICM.

With a large enough sample, we can also reverse the question: rather than study the influence of the
clusters on radio galaxies, we investigate if one can use radio galaxies to search for tell-tale
signs of nearby clusters. Previous surveys \citep{obrien16,paterno-mahler17} have used bent radio
sources to find clusters. A cluster was also discovered using a bent radio galaxy discovered in RGZ
\citep{banfield16}. We will investigate in this paper how bent radio galaxies can be used as a
statistical indicator of the galaxy overdensities around them.
 
In Section~\ref{sec:data}, we present the catalogs from which we drew our data. In
Section~\ref{sec:sample}, we describe our sample selection process. In Section~\ref{sec:qualifying},
we discuss sources and levels of uncertainty in our bending angle measurements, and how we correct
for that. In Section~\ref{sec:results}, we report our specific findings (summarized in
Section~\ref{sec:summary}). In Section~\ref{sec:discussion}, we discuss the physical interpretation
of our results and potential directions for follow-up research.

Throughout this work we assume a cosmology of $H_0=68~\textrm{km~s}^{-1}~\textrm{Mpc}^{-1}$,
$\Omega_\Lambda=0.685$, and $\Omega_m=0.315$ \citep{planck14}.

\section{Data} \label{sec:data}

The RGZ catalog is comprised of radio sources that citizen scientists have matched to infrared host
galaxies. The main value of RGZ comes from the matching of associated and possibly disconnected
radio source components, as well as the cross-identification of host galaxies. The bulk of the radio
data in the RGZ catalog comes from the Faint Images of the Radio Sky at Twenty Centimeters survey
\citep[FIRST;][]{becker95,white97}, and we combine radio galaxy classifications generated by the
citizen scientists with additional data derived from the radio contours from FIRST, which begin at a
level 4 times the local noise. The RGZ catalog also contains a sample of radio sources from the
Australia Telescope Large Area Survey \citep[ATLAS;][]{franzen15}, but those sources are not
included in this paper.

The host galaxies to which the radio sources are matched are identified on W1 band ($3.6~\mu$m)
images from the \emph{Wide-Field Infrared Survey Explorer} \citep[WISE;][]{wright10}. Each source
may or may not have an IR identification. Those that do have an IR counterpart identified by RGZ
consensus usually have a counterpart in the WISE catalog, but those with insufficient
signal-to-noise for inclusion in that catalog may not.

\begin{figure}
\centering
\includegraphics[width=0.45\textwidth]{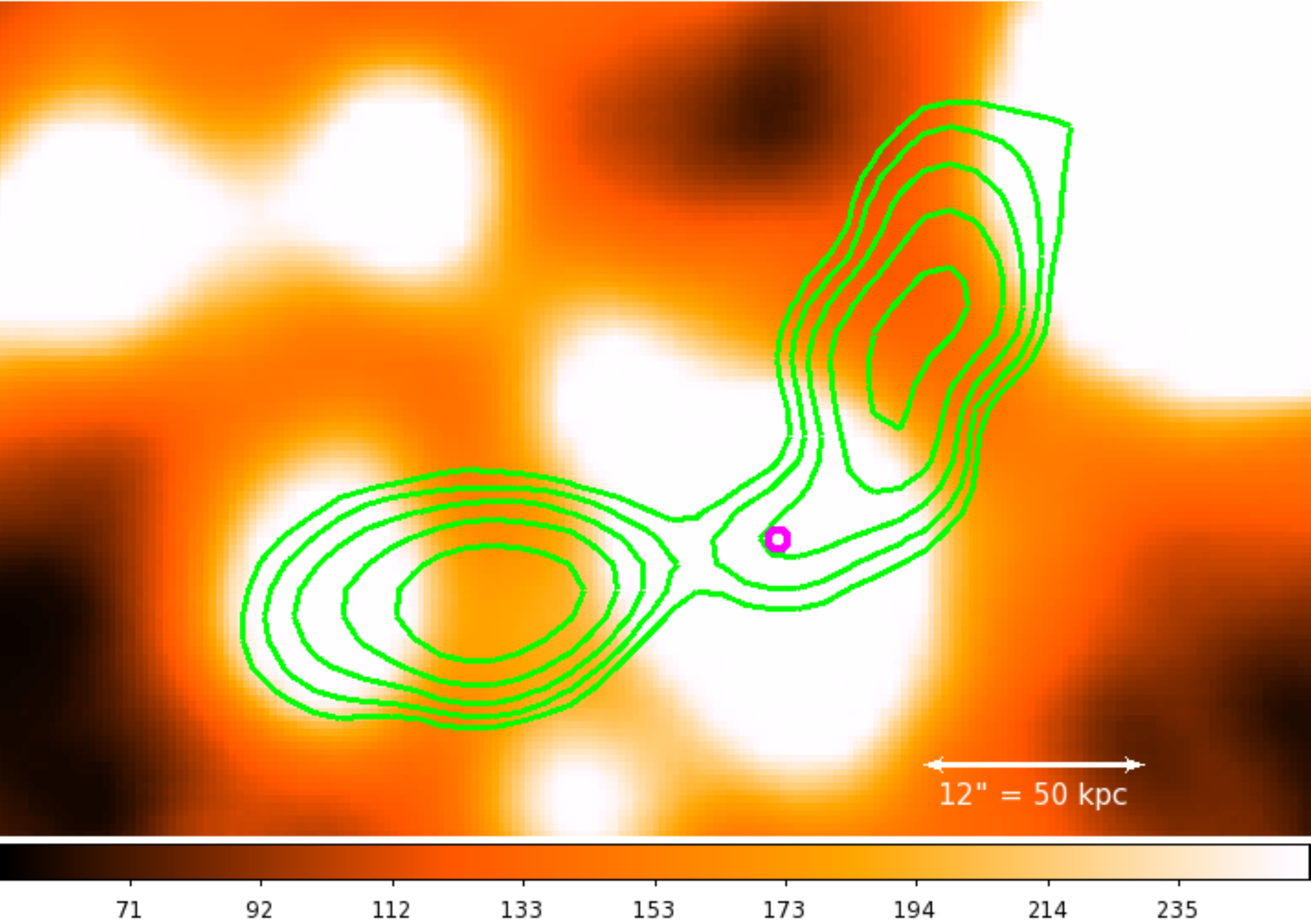}
\caption{Example source RGZ J080641.4+494629. This image has been cropped from the full $3'\times3'$
image shown to RGZ users. Background image is the IR field from WISE; contours are derived from
FIRST radio data. Consensus host position is marked with a circle, and we cross-matched it to galaxy
SDSS J080641.39+494628.4 at redshift $z=0.2447$; this marker is not visible in the image shown to
users. The IR emission extending to the northeast of the host is a foreground star blended in WISE.}
\label{fig:rgz}
\end{figure}

An example of the source as presented to RGZ users is shown in Figure~\ref{fig:rgz}. Radio contours
from FIRST are displayed on top of the IR field from WISE. The RGZ user also can toggle between
FIRST contours and a semi-transparent radio intensity map. In this example, the consensus host
galaxy in the IR image is located in the bridge of radio emission connecting the two outer lobes.

For this paper, sources with IR counterparts were cross-matched to the Sloan Digital Sky Survey DR13
catalog \citep[SDSS;][]{albareti17}, providing redshift measurements and improved angular resolution
over WISE. We performed the cross-match to the \texttt{PhotoPrimary} table in SDSS using a 3~arcsec
search radius. For sources that SDSS classified as extended based on the photometry, we queried the
\texttt{Photoz} table for photometric redshifts (present for 98\% of our sample; see also
\citealt{beck16}). We furthered queried the \texttt{SpecObj} table to get spectroscopic redshifts
where available (present for 40\% of our sample). We gave preference to the SDSS spectroscopic
redshift where available; otherwise, we used the SDSS photometric redshift. The photometric
redshifts have larger errors and the distribution peaks at a higher redshift than the distribution
of spectroscopic redshifts in our data set, suggesting that sources with only photometric redshifts
will be labeled as background sources more often; however, this should not systematically bias our
analysis.

We wrote an automated algorithm to identify double and triple radio sources from the radio contour
information and consensus user classifications. A double source was defined as either having two
disjoint radio components or a single radio component with two radio peaks. A triple source was
likewise defined as either having three disjoint radio components or a single radio component with
three radio peaks. Additionally for triple sources, one of the radio peaks was constrained to be
coincident with the SDSS position, i.e. the SDSS position must lie within the innermost contour
containing one of the peaks.

\begin{figure}
\centering
\includegraphics[width=0.45\textwidth]{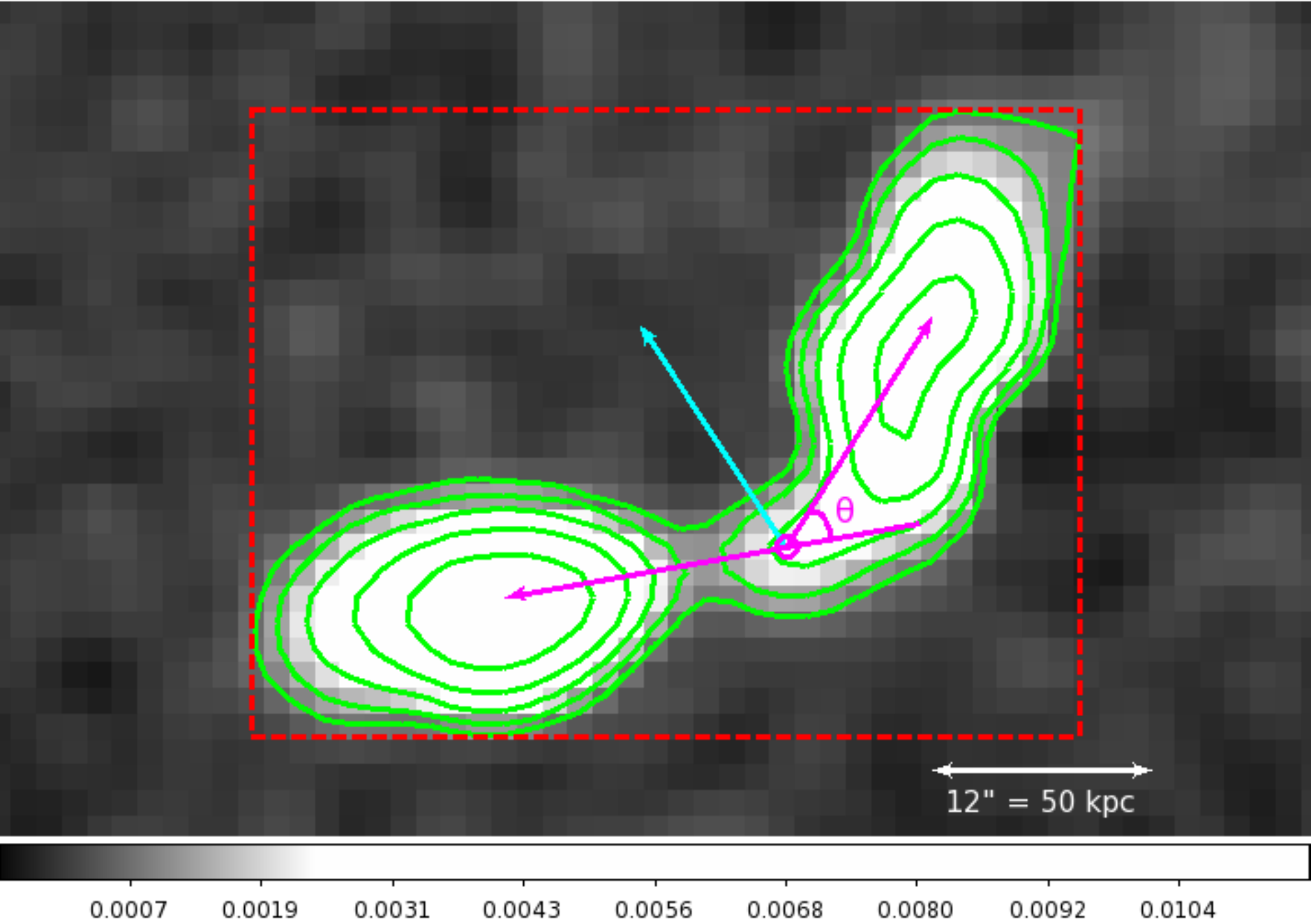}
\caption{FIRST image of the example source in Figure~\ref{fig:rgz}. The magenta circle is the
position of the host galaxy from SDSS; the magenta arrows terminate in the two radio peaks and are
used to measure the bending angle; the cyan arrow is the angle bisector of the opening angle; the
green contours are the radio contour levels; and the red dashed rectangle marks the bounding box
used to measure the angular size. This source has a bending angle of $\theta=47\degr$ and angular
size of 0.93~arcmin. At redshift $z=0.2447$, this corresponds to 220~kpc.}
\label{fig:example}
\end{figure}

Radio peaks were derived from the FIRST images as part of the RGZ pipeline. Within each innermost
contour in the FIRST image, the pixel with the greatest flux was defined as a peak. In this way, a
single radio component can have multiple peaks if there is substructure within the radio emission.
The majority of sources in RGZ contain only a single radio peak; after removing those, as well as
sources with more than three radio components (approximately 3\% of RGZ), we identified an initial
set of 6207 radio sources (4957 doubles and 1250 triples) for which we could measure the bending
angle. We used the SDSS position to calculate the bending angle by drawing lines from the host
through the two radio peaks (excluding the one coincident with the host in a triple source). The
bending angle is then defined as the deviation of the two lines from co-linear (e.g. a perfectly
straight source has bending angle $0\degr$). See Figure~\ref{fig:example} for an annotated example.

We used the updated version of the \cite{wen12} cluster catalog \citep[][hereafter WH15]{wen15} for
the sample of galaxy clusters. This catalog contains 158,103 clusters obtained from SDSS covering a
range of redshifts from 0.02 to 0.8 and includes optically-derived values for their richness
($R_{L*,500}$) and radius ($r_{500}$). It also includes a method of calculating an optical mass
proxy from the cluster richness, which WH15 derived by fitting to the X-ray- and SZ-derived masses
for a subset of the catalog,
\begin{equation} \label{eq:m500}
\log M_{500} = 1.08\log R_{L*,500} - 1.37,
\end{equation}
where $M_{500}$ is the cluster mass within $r_{500}$ and $R_{L*,500}$ is the cluster richness,
defined as the luminosity within $r_{500}$ normalized by $L^*$.

\cite{wen12} identified clusters in SDSS using a friends-of-friends algorithm applied to luminous
galaxies, $M_r^e(z) \leq -21$, where $M_r^e(z)$ is the evolution-corrected absolute $r$-band
magnitude. In each cluster candidate, the galaxy with the greatest number of friends was taken as
the temporary center of the cluster, then the most luminous galaxy within 0.5~Mpc and $\pm
0.04(1+z)$ of the temporary center was defined as the BCG. Finally, the position of the BCG was
defined to be the center of the cluster. \cite{wen12} also applied a minimum richness threshold of
$R_{L*,200} \geq 12$, which corresponds to $M_{200} \sim 0.6\times 10^{14}~M_\sun$.

We matched a radio source to a cluster by finding the nearest cluster located within a projected
distance of 15~Mpc from the host galaxy, calculated at the redshift of the galaxy, and within a
redshift range of $|\Delta z|<0.04$, where $\Delta z$ is defined as
\begin{equation} \label{eq:dz}
\Delta z = \frac{z_{gal}-z_{cluster}}{1+z_{gal}}.
\end{equation}

We chose the 15~Mpc projected distance as being typical of the distance between clusters, so any
galaxy associated with a cluster should be matched within that distance. At higher redshifts the
completeness of WH15 decreases, but this fiducial separation is much larger than the separation
within which we expect to observe an effect due to cluster proximity.

\begin{figure}
\centering
\includegraphics[width=0.49\textwidth]{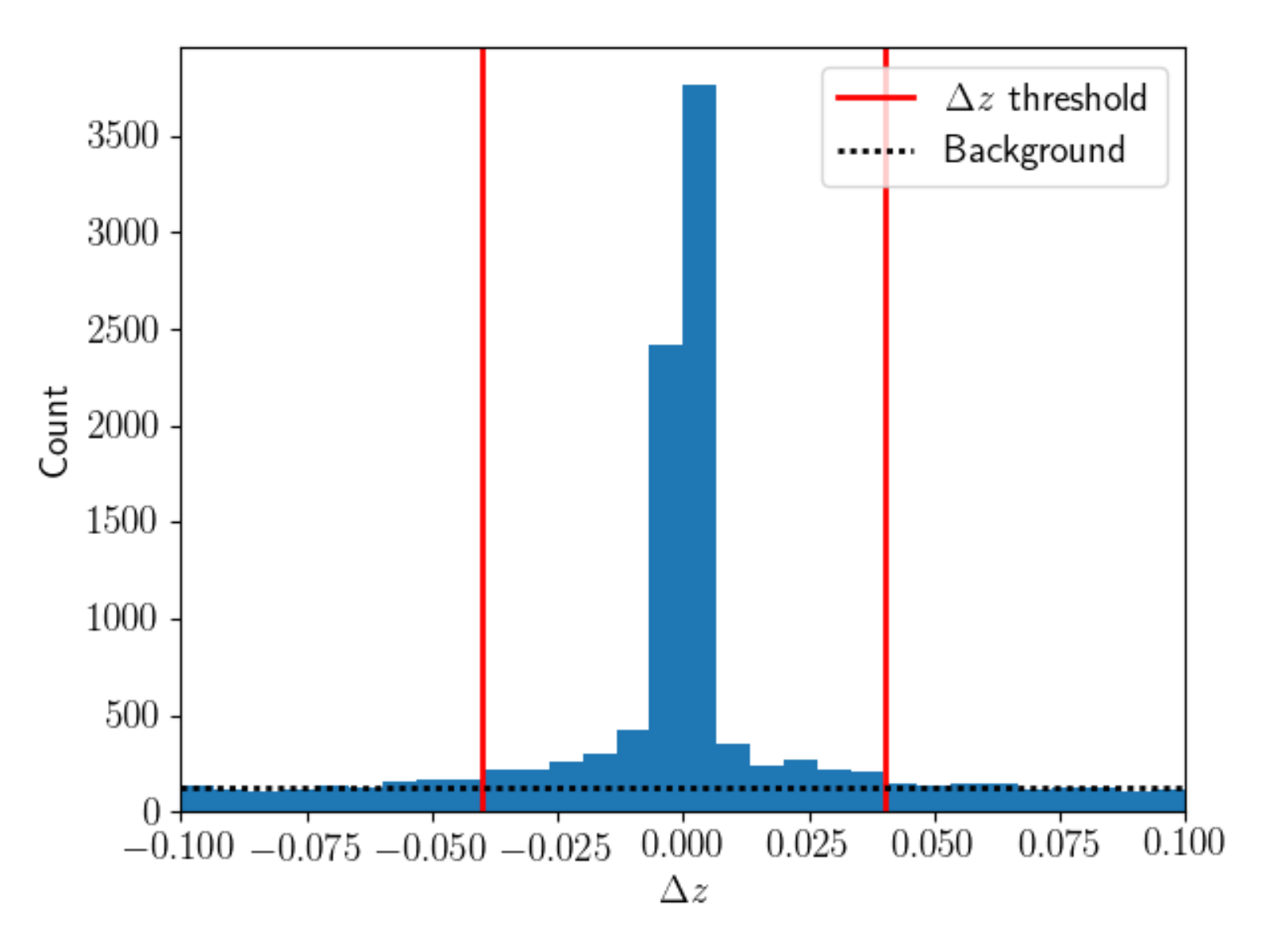}
\caption{The distribution of $\Delta z$ (defined in Equation~\ref{eq:dz}) in the RGZ-WH15 15~Mpc
positional cross-match, plotted over the range $-0.1$ to $+0.1$. The solid vertical lines are the
$\Delta z$ threshold used for matching, and the dotted horizontal line is the estimated contribution
due to background sources.}
\label{fig:dz}
\end{figure}

The choice of a redshift range within which to consider a radio galaxy matched to a cluster
represented a trade-off between including as many true cluster radio galaxies as possible, and
simultaneously minimizing the number of contaminating radio galaxies. The $\Delta z$ distribution is
shown in Figure~\ref{fig:dz}, within the projected 15~Mpc threshold. We chose a redshift threshold
of $|\Delta z|<0.04$ because there was a clear excess of cluster sources within that range. This
happens to be the same redshift range criterion used in the generation of the WH15 catalog. The
threshold is marked in Figure~\ref{fig:dz} with solid vertical lines. This unavoidably results in
accepting some small number of radio galaxies, estimated at the level of the dotted horizontal line,
which are actually as much as a comoving $\pm100$~Mpc from their matched cluster.

We discuss the contamination and completeness due to these thresholds in Section~\ref{sec:contam};
however, any contribution from contaminating sources (including ones with incorrect host
identifications) will be uncorrelated with the cluster parameters we study, so it will merely serve
to lower the reported significance level of our results. With both projected separation and redshift
cuts, we have 4257 double sources and 1096 triple sources in the matched sample. In other words,
87\% of our sample is located within a projected 15~Mpc of a cluster.

We calculate the orientation angle of the radio source relative to the cluster center by measuring
the difference between the angle bisector of the opening angle and the radial vector to the cluster
center. This value ranges between $0\degr$ for sources opening parallel to the radial vector (i.e.
bent away from the cluster center) and $180\degr$ for sources opening anti-parallel to the radial
vector (i.e. bent towards the cluster center).

\section{Sample selection} \label{sec:sample}

We restricted the sample to sources with at least a 0.65 consensus in the radio component
classification from RGZ to remove sources with ambiguous structure, as well as non-physical
associations between radio lobes and coincident galaxies. A minimum statistical reliability of 80\%
was found for the 0.65 consensus limit for RGZ DR1 (O. I. Wong et al., in prep.). This consensus
value includes the use of classifier weighting and was calibrated to the classifications performed
performed by the RGZ science team.

No reliability tests were performed for the IR host identifications, and we did not apply a
consensus threshold for those identifications here. As described in \cite{banfield15}, the host
galaxy of a radio source is identified by a kernel density estimator (KDE) method based on the
clustering of clicks from the RGZ participants. While spectral indices are not available to verify
the host galaxy selection, the ambiguity of the host galaxy candidates is reduced via the
combination of 1) the low-redshift volume; 2) the radio galaxy angular size selection criteria; and
3) the cross-identification of WISE and SDSS observations. In Section~\ref{sec:correction}, we find
that double sources follow the same trends as triple sources, suggesting that host identifications
even in the absence of a radio core are statistically as reliable as those with radio cores; and in
Section~\ref{sec:discussion}, we find that our results are generally consistent with previous
studies which used more manual verification of a few tens of radio sources
\citep[e.g.][]{blanton01,rodman19}. Such consistencies suggest that our results are unlikely to be
contaminated by a significant number of misidentified host galaxies.

\begin{table}
\centering
\begin{tabular}{ccccc}
Cut & Double & Triple & Unmatched & Total \\
\hline
RGZ catalog & N/A & N/A & N/A & 118,090 \\
Extended & 4257 & 1096 & 854 & 6207 \\
Reliability & 3202 & 477 & 577 & 4256 \\
$\theta_{corr}$ & 3235 & 488 & 581 & 4304 \\
\hline
\end{tabular}
\caption{Summary of the sample size after each cut in the sample selection process. ``Extended''
refers to extended sources found using the classification algorithm described in
Section~\ref{sec:data}, ``reliability'' refers to the multiple reliability cuts described in
Section~\ref{sec:sample}, and ``$\theta_{corr}$'' refers to reapplying the bending angle threshold
using the bending angle correction described in Section~\ref{sec:correction}.}
\label{tab:cuts}
\end{table}
 
Given the 3~arcmin size of RGZ subject fields, we applied a radio source size threshold of
1.5~arcmin to ensure that sources are wholly contained within the field. We use the RGZ catalog's
measure of largest angular size, which is defined as the diagonal across the bounding box containing
the outer ($4\sigma$) contour level of the radio emission. Applying a maximum angular size will
preferentially retain large sources at high redshift, which tend to be more luminous
\citep{kapahi85}; we do not expect this to bias our results. We also removed any sources that
intersect the edge of the subject field presented to RGZ users.

We imposed a bending angle maximum of $135\degr$, because above this, a brief visual inspection
revealed that such highly bent angles are probably due to misidentification of the host far from the
major axis of radio emission. This excludes the small fraction of true narrow-angle tail sources
which may be in the sample. With all of these requirements, the matched sample was reduced to 3203
double sources and 477 triple sources.

\begin{figure}
\centering
\includegraphics[width=0.49\textwidth]{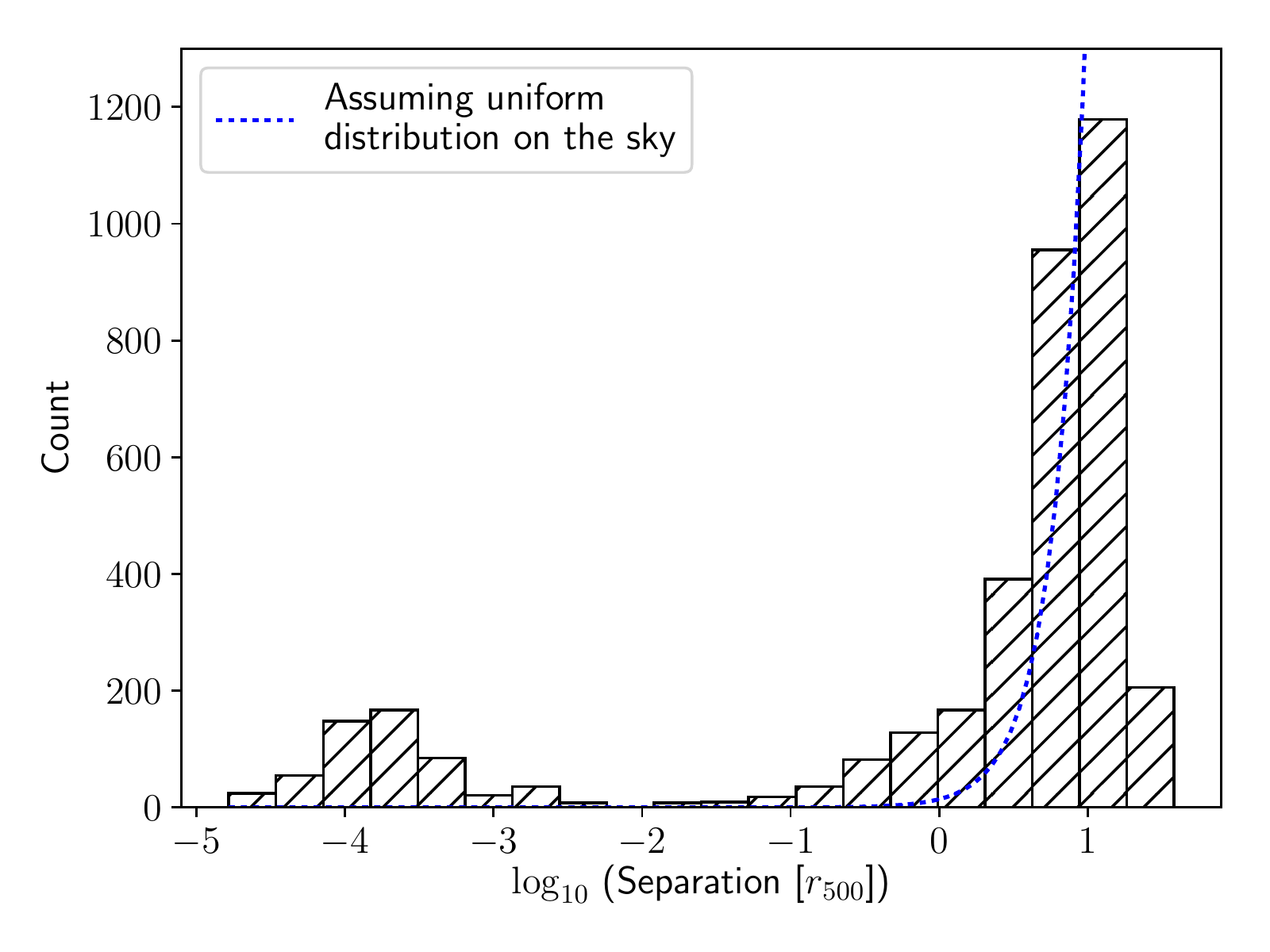}
\caption{Histogram of normalized separations in the sample. The smaller peak near $10^{-4}$
corresponds to the BCGs of the clusters, which are defined in WH15 as being at the center. The
dotted line represents the distribution if the sources were uniformly distributed on the sky within
$18.8~r_{500}$ of the nearest cluster, which was chosen as the distance that encompasses 95\% of the
sources in our matched sample.}
\label{fig:sep_hist}
\end{figure}

In Section~\ref{sec:correction}, we define a ``corrected bending angle,'' $\theta_{corr}$, based on
the matched sample of 3680 sources, and update the bending angle threshold of $135\degr$ to refer to
the corrected bending angle. This increases our final sample to 3742 double sources and 562 triple
sources, of which we match 3235 double sources and 488 triple sources to clusters within a projected
15~Mpc. The sample size after each cut is summarized in Table~\ref{tab:cuts}. The final sample of
4304 sources are included as supplemental material on the journal website, along with all of the
derived quantities used within this work; see the Appendix for details.

\subsection{Spatial distribution of radio galaxies around clusters} \label{sec:sample_dist}

For each source in the matched sample, we take its projected distance from the center of the cluster
(now calculated at the redshift of the cluster to be more physically meaningful, in contrast to the
matching algorithm in Section~\ref{sec:data}) and normalize it by the $r_{500}$ of the cluster, as
given in the WH15 catalog. The median $r_{500}$ in WH15 is 0.65~Mpc. The histogram of normalized
separations is shown in Figure~\ref{fig:sep_hist}. The distribution is bimodal, with one peak around
$10^{-4}~r_{500}$ and the other around $10~r_{500}$. The first peak is due to the WH15 definition of
the cluster center, which is defined as the position of the BCG; because WH15 used Data Releases 8
and 9 of SDSS, whereas RGZ uses Data Release 13, there is often a small positional offset (less than
1~arcsec) between the WH15 and RGZ position of the host galaxy. Given this, we treat all sources
within $0.01~r_{500}$ of the cluster center as being the BCG of their respective cluster and group
them together in our analysis. This corresponds to a physical separation of only 7~kpc for the
median cluster in our sample, well within the size of a galaxy.

\begin{figure}
\centering
\includegraphics[width=0.49\textwidth]{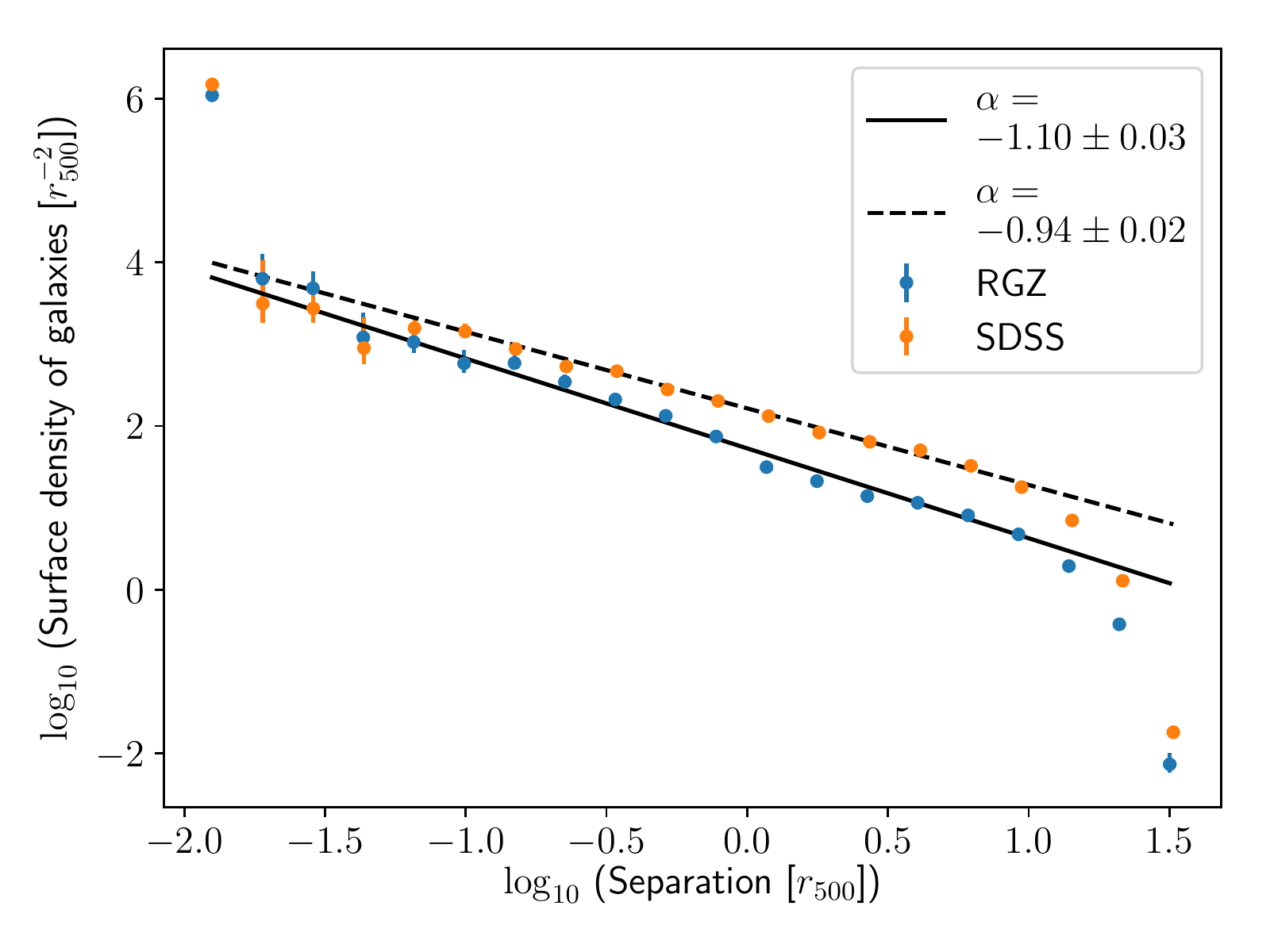}
\caption{Surface density (count per square $r_{500}$) as a function of normalized separation. The
best fit power law (with index $\alpha$) is shown. The sample of radio galaxies from RGZ is shown in
blue with a solid power law fit. The matched sample of optical galaxies from SDSS is shown in orange
with a dashed power law fit.}
\label{fig:density}
\end{figure}

When plotting the number of sources matched to a cluster per unit area around it, the surface
density scales as a power law. We perform a linear fit to the data in loglog space between 0.01 and
$10~r_{500}$ and calculate that the best fitting power law has index $\alpha = -1.10\pm 0.03$, shown
as the blue points and solid line in Figure~\ref{fig:density}. We compared this result to a sample
of 14,214 optically selected galaxies from SDSS. We selected the SDSS sample by drawing random RA,
Dec., and redshift values following the distribution of the RGZ sample and taking the nearest
corresponding galaxy in SDSS. Using the same cluster-matching algorithm, we matched 13,182 (93\%) of
the SDSS galaxies to WH15. Figure~\ref{fig:density} shows the surface density distribution for the
optical galaxies in orange, for which the best fitting power law (dashed line) has index $\alpha =
-0.94\pm 0.02$.

Both radio and optical distributions follow an approximately $1/r$ trend, consistent with an
isothermal distribution within the cluster. Since a background population would have a flat profile,
these distributions are likely dominated by true cluster galaxies. However, the power law slope is
slightly but statistically significantly steeper for cluster radio galaxies. $271/2209=(12.3\pm
1.0)\%$ of non-BCG radio sources within $10~r_{500}$ of the nearest cluster are within the innermost
$r_{500}$ of that cluster, compared to $584/8823=(6.6\pm 0.3)\%$ of optically-identified galaxies
within $10~r_{500}$. This indicates that extended radio galaxies are a biased population (consistent
with \citealt{overzier03} and references therein). We believe the cut-off in the power law beyond
$10~r_{500}$ is due to a combination of the typical size of a galaxy cluster and when galaxies begin
matching to another nearby cluster, although quantifying that effect would require a more detailed
analysis.

In addition, we find that BCGs are at least 2.5 times as likely to have extended radio emission as
the general population of cluster galaxies: $546/2755=(19.8\pm 1.2)\%$ of the RGZ matches within
$10~r_{500}$ are BCGs, compared to only $743/9566=(7.8\pm 0.4)\%$ of the SDSS matches within the
same separation. This is consistent with previous findings \citep{best07}.

\subsection{Contamination} \label{sec:contam}

Figure~\ref{fig:dz} shows the distribution of $\Delta z$ in the RGZ-WH15 cross-match. We use this
distribution to estimate the contamination and completeness of our sample. Given our redshift
matching criterion of $|\Delta z| < 0.04$, we take the mean number of matches in each $|\Delta z|$
bin between $0.04$ and $0.1$ as the number of contaminating sources per bin. Between $|\Delta z| =
0$ and 0.04, 21.5\% of the matched sources can be attributed to this contamination. Likewise, we
take the number of sources that are above this contamination level in bins with $|\Delta z|$ between
0.04 and 0.1, where the entire sample is likely from contamination, as the number of sources that
should be included in our sample but are removed by the redshift threshold. 1.8\% of the
cross-matched sources are removed this way, leading to a sample that is at least 98.2\% complete.

\begin{figure}
\centering
\includegraphics[width=0.49\textwidth]{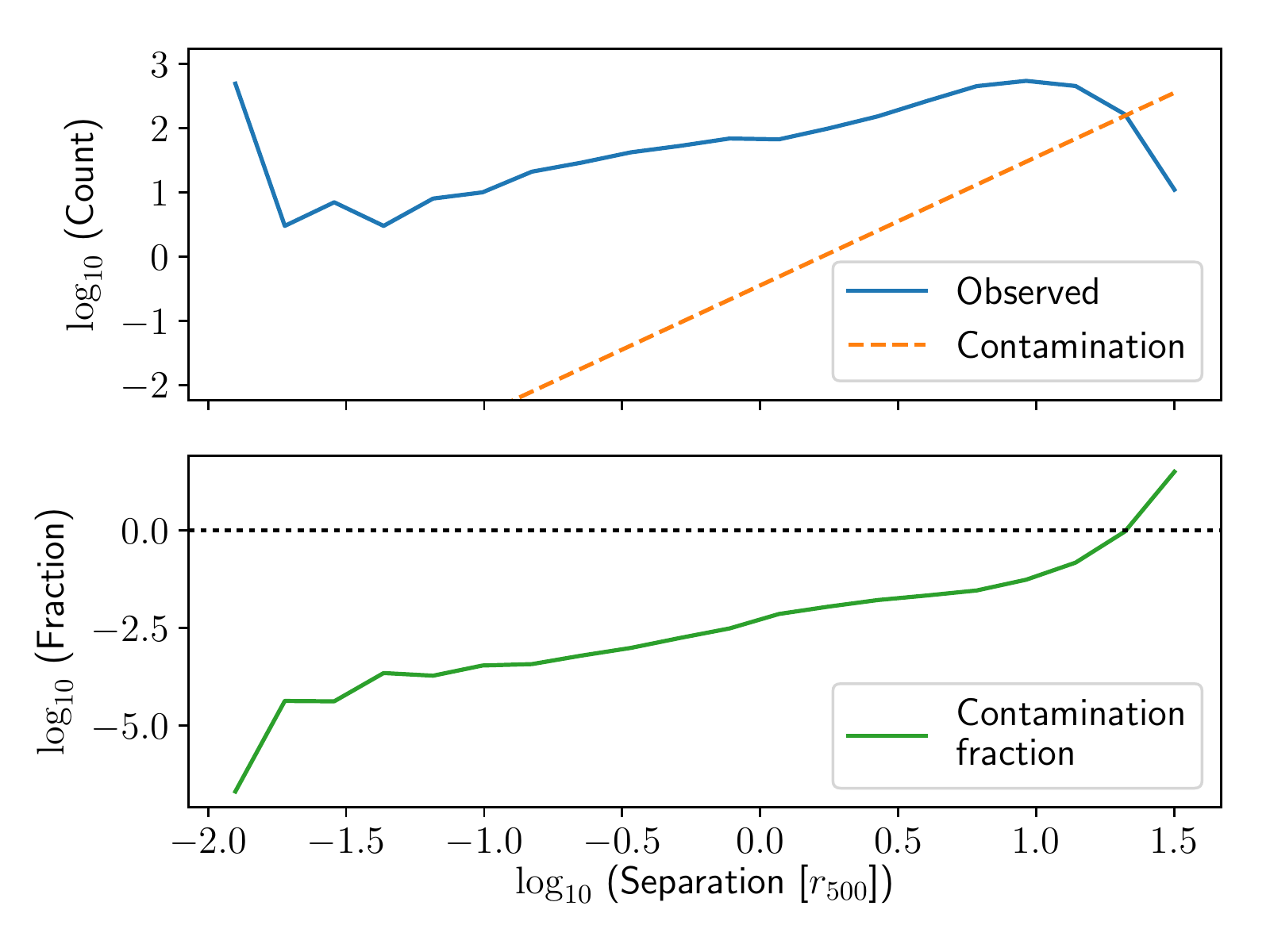}
\caption{\emph{(Top)} The solid line is the same distribution as the histogram of normalized
separations in Figure~\ref{fig:sep_hist}, except with all bins within $0.01~r_{500}$ combined. The
dashed line is the estimated number of contaminating sources in each bin, assuming a uniformly
distributed contaminating population. \emph{(Bottom)} The fractional contamination in each bin. The
dotted line marks 100\% contamination, which occurs beyond $21.1~r_{500}$ ($\sim 13.7$~Mpc).}
\label{fig:contam}
\end{figure}

We estimate the effect of contamination as a function of cluster separation. Assuming that
contaminating sources are uniformly distributed on the projected sky, we calculate the number and
fraction of expected contaminating sources at each cluster separation. The top plot of
Figure~\ref{fig:contam} shows the distribution of sources in our sample (solid; the same
distribution as Figure~\ref{fig:sep_hist}) and the number of contaminating sources in each bin
(dashed; note that this is not the cumulative contamination). In our analysis we focus on sources
within $1.5~r_{500}$ of their nearest cluster, which also corresponds to approximately the virial
radius; the contamination in the bin at this separation is only 0.01\%. A more detailed
contamination analysis would include the effect of galaxies at high separations matching to other
neighboring clusters, which we predict contributes to the drop-off in source count that occurs
beyond $10~r_{500}$. However, we can safely say that the effect of contamination in the range of
separations in which we are interested is minimal.

\section{Qualifying the bending angle} \label{sec:qualifying}

\subsection{Bending angle uncertainties} \label{sec:uncertainty}

Our definition of bending angle is based on the angle between the host galaxy and the peaks in the
radio emission. This is a well-defined and easily measurable, but still arbitrary definition. The
first thing we do is assess the measurement uncertainties in bending angle that arise from this
definition and the properties of our data sample.
 
There are four known sources of potential error in our measurement. The first is radio component
classification error, which occurs when the radio source could consist of a chance superposition of
unrelated radio components. We attempted to minimize this with the 0.65 consensus threshold in the
sample selection stage.

The second is astrometry error of the optical host, including a very small contribution from host
misidentifications. The astrometry error typically induces less than a tenth of a degree of bending,
given the angular size of the sources and the sub-arcsec precision in the positions of the host
galaxies in SDSS. Host misidentification could also contribute an error due to a spurious match
between a misidentified host and a cluster unrelated to the radio source, but as discussed in the
previous section, contamination by these foreground or background sources will have a negligible
effect on our results.

The third is projection effects; while the radio sources exist in three-dimensional space, we can
only measure the bending angle projected onto the sky. Because sources are oriented isotropically in
space, the observed bending angle for a given source could be either increased or decreased by
projection; for a large sample however, the median observed bending angle remains consistent with
the true bending angle for all but the most bent sources ($\theta_{true}\gtrsim 60\degr$). Even
then, projection will on average tend to reduce the observed bending angle, suggesting that any
trends we observe with respect to bending angle are merely lower bounds.
 
\begin{figure}
\centering
\includegraphics[width=0.49\textwidth]{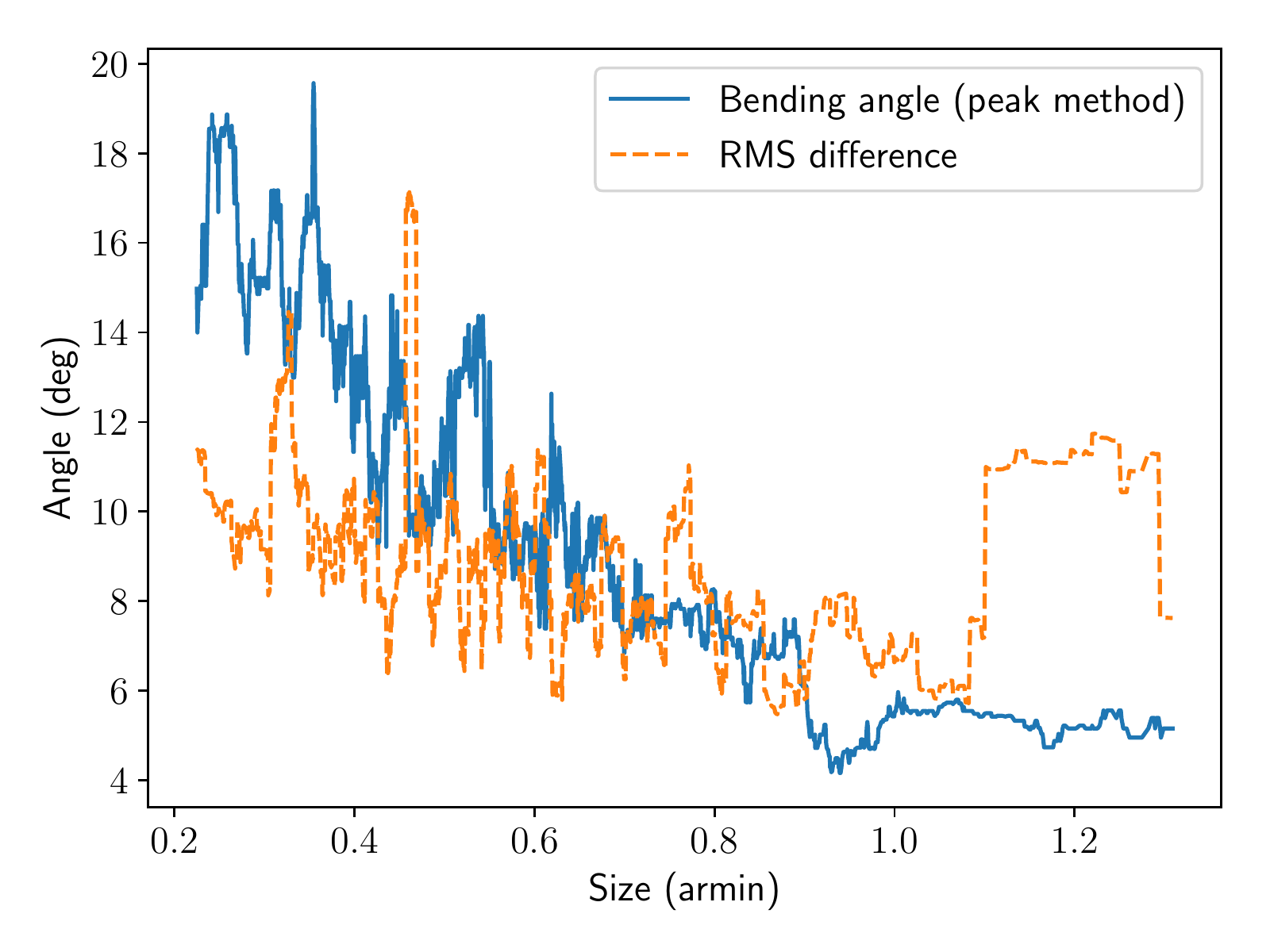}
\caption{Comparison between the bending angle and bending angle error due to morphology uncertainty.
The solid line is the measured bending angle as a function of source angular size, smoothed using a
window size of 100 sources. The dashed line is the running root-mean-square difference between
bending angle calculated through the radio peaks and bending angle calculated through the most
distant point on the contours. The RMS difference is not used in the correction; it simply
demonstrates the magnitude of the morphology effect on bending. The large increase in RMS for
sources larger than 1.1~arcmin is due to a small number of outliers, but does not change our
results.}
\label{fig:bending_rmsd}
\end{figure}
 
The fourth and largest source of error is uncertainty due to structure in the radio emission. We
estimate the amount of error introduced by this uncertainty by comparing the measurement derived
from our definition of bending angle to an alternative measurement of the angle between the host
galaxy and the most distant point on the lowest contour on each side that does not contain the host.
We calculate the running root-mean-square difference between the two measures of bending angle,
shown in Figure~\ref{fig:bending_rmsd}. This follows an approximately linear relationship with
angular size, decreasing from an estimated uncertainty of $11\degr$ for sources of size 0.2~arcmin
down to $6\degr$ for sources of size 1.1~arcmin. The divergence from a linear relationship for
sources larger than 1.1~arcmin is due to a small number of outliers, but does not change our
results.

From geometric arguments, the bending angle error is proportional to the error in the position of
the host and inversely proportional to the angular extent of the source, because the angle will be
amplified if the radio peaks are closer to the host. Bending angle is a positive definite quantity,
so the errors are not symmetric; given the same ``true'' bending angle, a larger error will
statistically make the source look more bent, so small sources have their measured bending angle
amplified.

The observed bending angle as a function of angular size for sources larger than $\sim$0.4~arcmin is
very similar to the empirically determined bending angle uncertainty. In general, we argue that this
similarity in magnitude and trend shows that the relationship between bending angle and angular size
is predominantly due to irreducible measurement uncertainties due to source structure, and is not
reflective of a physical trend. For smaller sources, the error gets compounded by the fact that
structure becomes harder to resolve in the FIRST images.

\subsection{Correcting for the dependence on angular size} \label{sec:correction}

Given that the bending angle is dependent on the angular size of the source, due in part to inherent
uncertainty in our measurement of bending angle, we correct for this relationship prior to analyzing
the correlation of bending with other physical quantities. In particular, source size is correlated
with the separation between the source and the cluster center. We observe that sources beyond
$1.5~r_{500}$ are on average 24\% larger in physical extent than sources within $1.5~r_{500}$. (We
show in Section~\ref{sec:results_sep} that beyond $1.5~r_{500}$, the bending is statistically
unaffected by the cluster, hence the cutoff here.) While part of this correlation is physical
\citep[e.g.][]{machalski06}, selection effects such as redshift prevent us from further analyzing
this trend.

\begin{figure}
\centering
\includegraphics[width=0.49\textwidth]{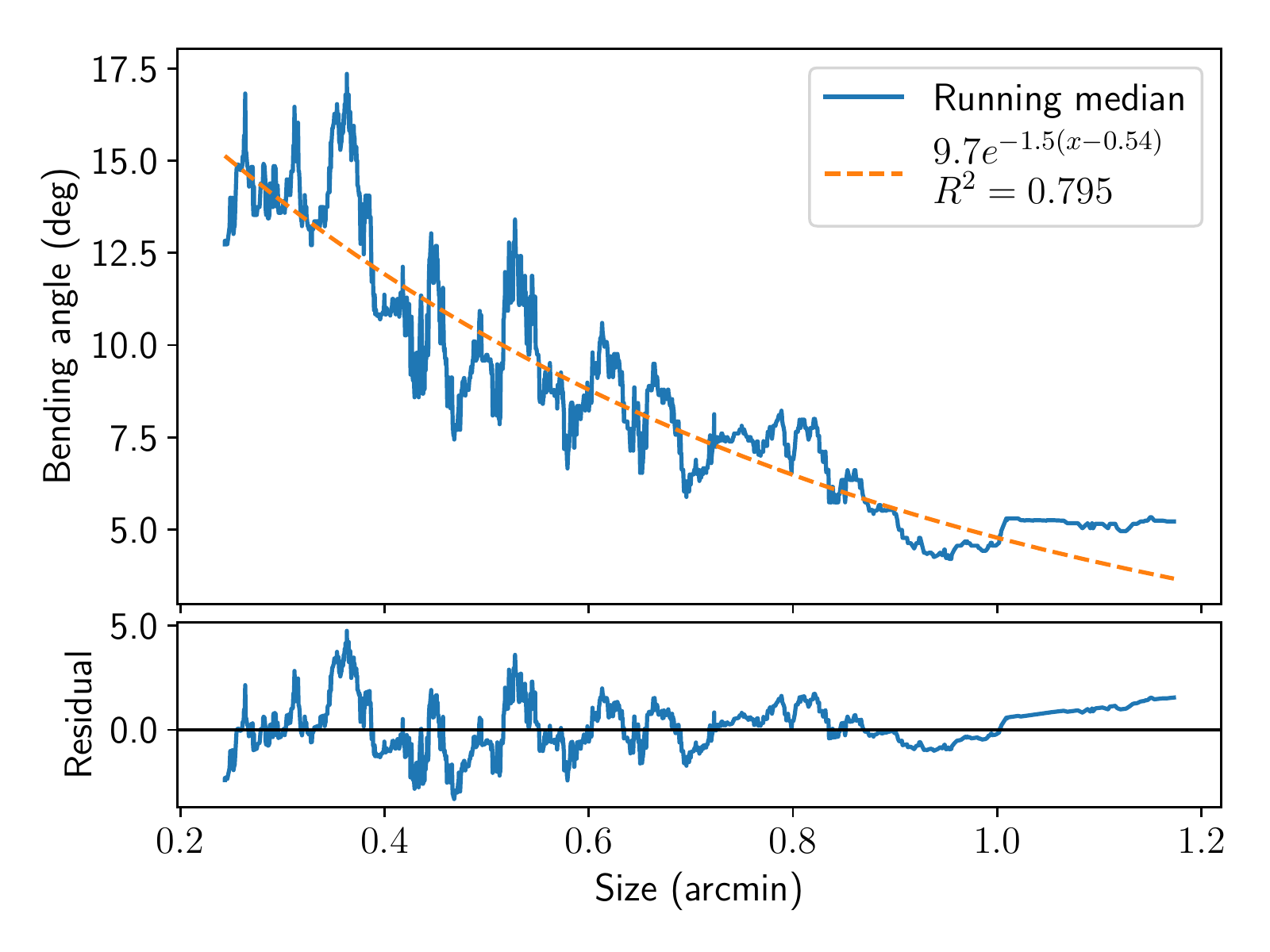}
\includegraphics[width=0.49\textwidth]{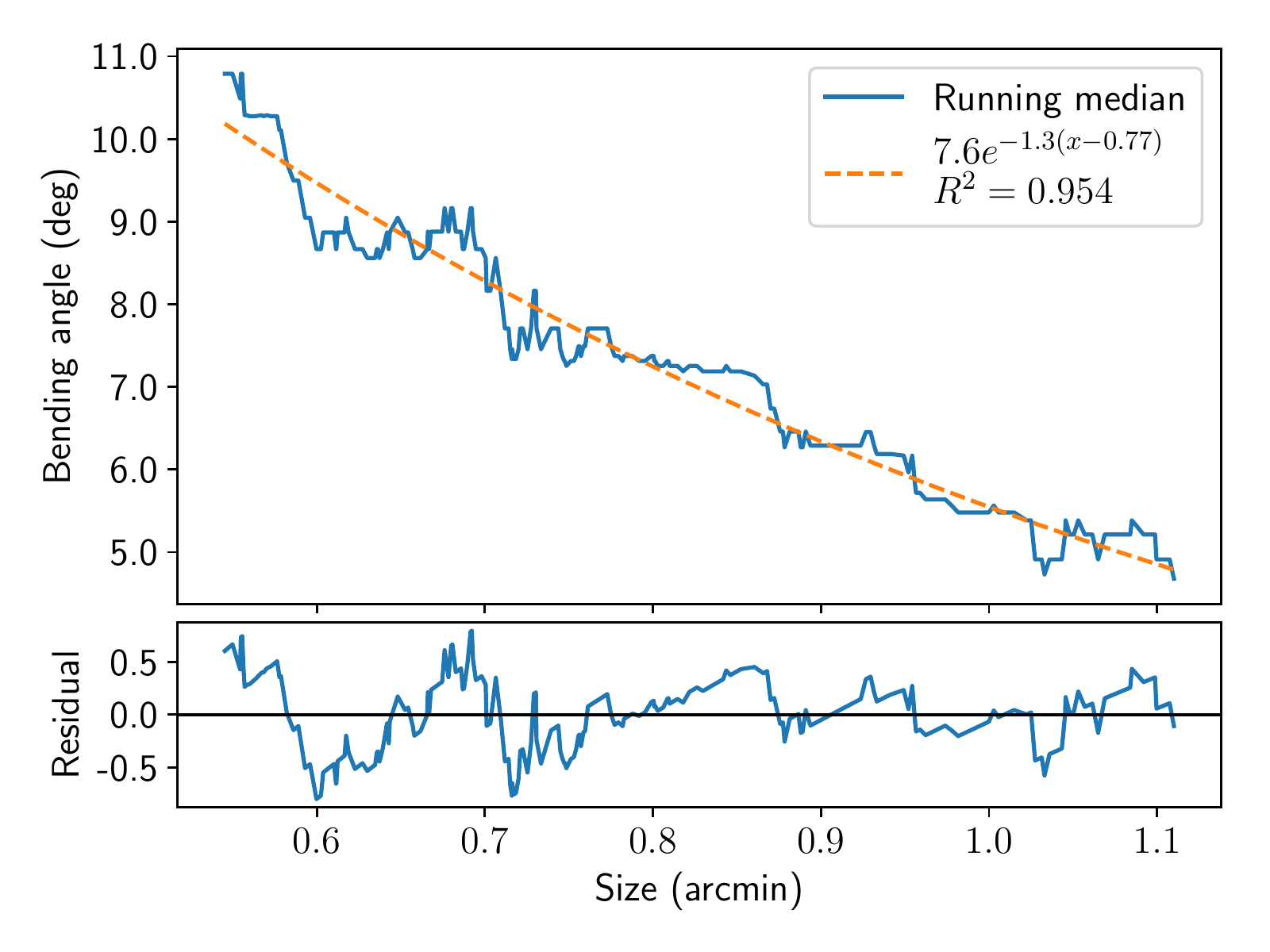}
\caption{Best fit and residuals for fitting the bending angle-angular size relationship.
\emph{(Top)} Double sources. \emph{(Bottom)} Triple sources.}
\label{fig:bending_correct}
\end{figure}

Because we are interested in the connection between bending and separation, among others, we must
remove this confounding factor. To make this correction, we take the sources that are beyond
$1.5~r_{500}$ and fit an exponential function to the median angular size-bending relation. We define
a ``corrected bending angle'' as the ratio of a source's measured bending angle to the predicted
bending angle given this relationship (fit shown in Figure~\ref{fig:bending_correct}). We do this
separately for double and triple sources for reasons explained in the next paragraph. For each
morphology, we normalize it such that a source of median size was unaffected by the correction.
After this correction, both the median and quartile distributions are independent of source size,
except for a minor deviation by the largest sources. We update the bending angle threshold of
$135\degr$ to apply to the corrected bending angle rather than the measured bending angle. This
results in our final sample of 3742 double sources and 562 triple sources (4304 total), of which
3235 double sources and 488 triple sources are matched to clusters within a projected 15~Mpc (3723
total).

\begin{figure}
\centering
\includegraphics[width=0.49\textwidth]{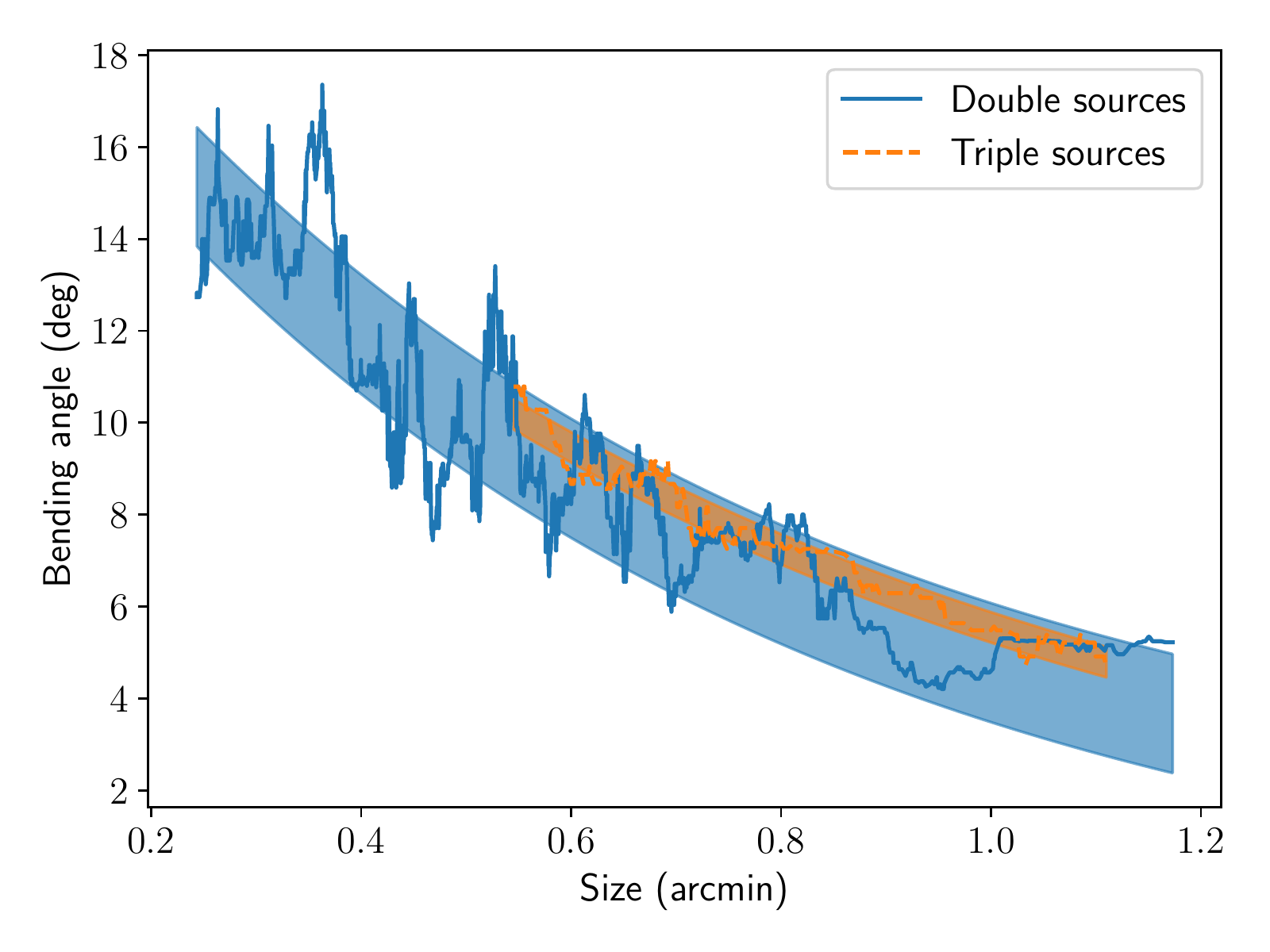}
\caption{Bending angle-angular size relationship for double (solid) and triple (dashed) sources in
our sample, as in Figure~\ref{fig:bending_correct}. The shaded regions are the best fit lines plus
or minus the RMS error in the fit.}
\label{fig:correction_comp}
\end{figure}

We corrected double and triple sources separately as a consistency test. Triple sources have a low
rate of false positive classifications, because they have radio emission at the position of the host
galaxy. However, there may exist confusion in the classification of double sources; for example,
there may be an unrelated infrared source in the WISE field equidistant between two regions of
potentially unrelated radio emission that gets mistaken as the host of a double source. If the host
identifications in our sample are predominantly real, we expect the size-angle trend to be the same
regardless of whether they are double or triple sources. A significant misidentification rate of
double sources would result in a different relationship between angular size and bending angle for
double and triple sources. Figure~\ref{fig:correction_comp} compares the size-angle trend for double
and triples sources, and we find that the trend for double sources is consistent with the trend for
triple sources. Therefore, we conclude that our population of double sources is dominated by correct
host identifications.

Bending angle is a positive definite quantity, and therefore has a contribution due to the
uncertainty alone. Using the corrected bending angle, we therefore define an ``excess bending
angle'' for each source as
\begin{equation} \label{eq:correction}
\Delta\theta = \sqrt{\theta_{corr}^2 - \theta_{med,outer}^2}
\end{equation}
where $\theta_{med,outer}=7.3\degr$ is the median corrected bending angle for sources matched to
clusters beyond $1.5~r_{500}$ (the \emph{outer region} defined in Section~\ref{sec:results_sep}).
This value is consistent with the irreducible uncertainty in measuring bending angle, which supports
our argument that sources beyond $1.5~r_{500}$ are outside the cluster's range of influence on
bending and are thus nominally straight. For the purposes of plotting, when
$\theta_{corr}<\theta_{med,outer}$ and $\Delta\theta$ becomes imaginary, we plot $-|\Delta\theta|$
instead; a negative excess by this definition is consistent with zero excess. More sophisticated
estimates of the most probable value of $\Delta\theta$ when $\theta\lesssim 20\degr$ would be
required if one were to compare the quantitative bending to specific physical models.

\section{Results} \label{sec:results}

We start by looking at the distribution of excess bending angle as a function of the galaxy's
distance from the cluster center, normalized by the cluster $r_{500}$. Then, we look at this
distribution as a function of the cluster $M_{500}$. We next combine these into a single
relationship, and look at this distribution as a function of ICM gas pressure. These three
quantities will serve as proxies for ICM density, galaxy velocity, and ram pressure, respectively.
We present an analysis on the distribution of radio tail orientations and claim it as a proxy for
the distribution of radio galaxy velocity vector orientations. We also discuss the frequency that
bent radio sources reside within clusters versus outside them, and investigate the local environment
around radio galaxies outside of clusters. Finally, we look at asymmetry in the radio tails. A
summary of the key results is presented at the end of this section.

\subsection{Excess bending angle versus normalized separation} \label{sec:results_sep}

Galaxy clusters have ICM density profiles that generally peak strongly towards the center
\citep{pratt02,newman13}, although there exist clusters with a dynamic or clumpy ICM where this
assumption may not hold (e.g. merging systems like the Bullet cluster). For a given galaxy velocity
with respect to the ICM, a denser environment should cause greater distortion of the radio jets.
Therefore, we expect to see a correlation between the bending of the radio source and its proximity
to the cluster center.

\begin{figure}
\centering
\includegraphics[width=0.49\textwidth]{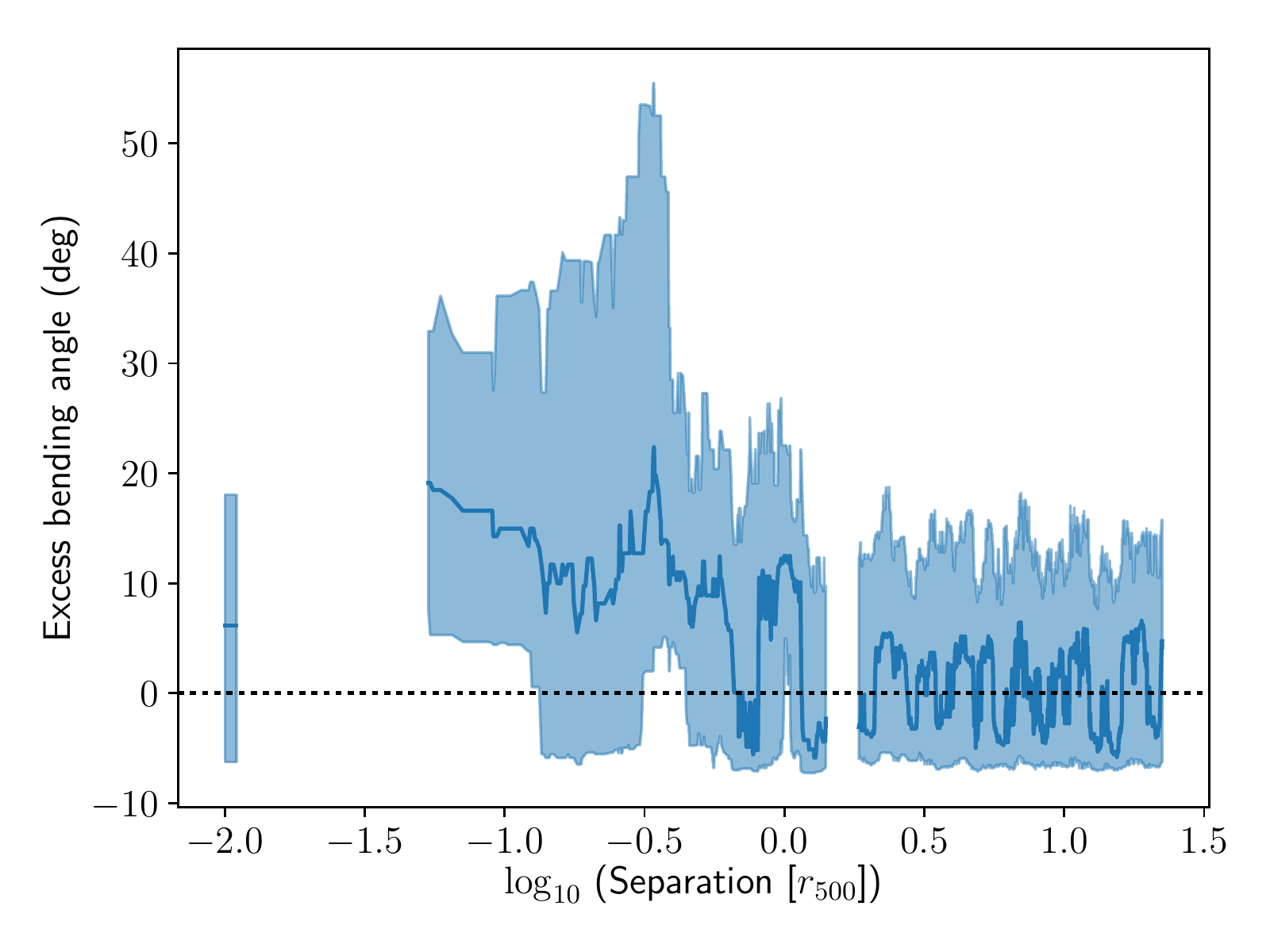}
\caption{Running median of excess bending angle ($\Delta\theta$) as a function of normalized
separation. The shaded region represents the range between the running first and third quartiles.
The running bending angles for the three regions we define (BCGs, \emph{inner}, and \emph{outer})
are calculated separately, denoted by the breaks in the plot. $\Delta\theta=0$, marked with the
dotted line, was defined in Equation~\ref{eq:correction} from the median corrected bending angle for
sources in the \emph{outer region}. Sources below the dotted line are consistent with straight.}
\label{fig:excess_vs_r}
\end{figure}

\begin{figure*}
\centering
\includegraphics[width=0.95\textwidth]{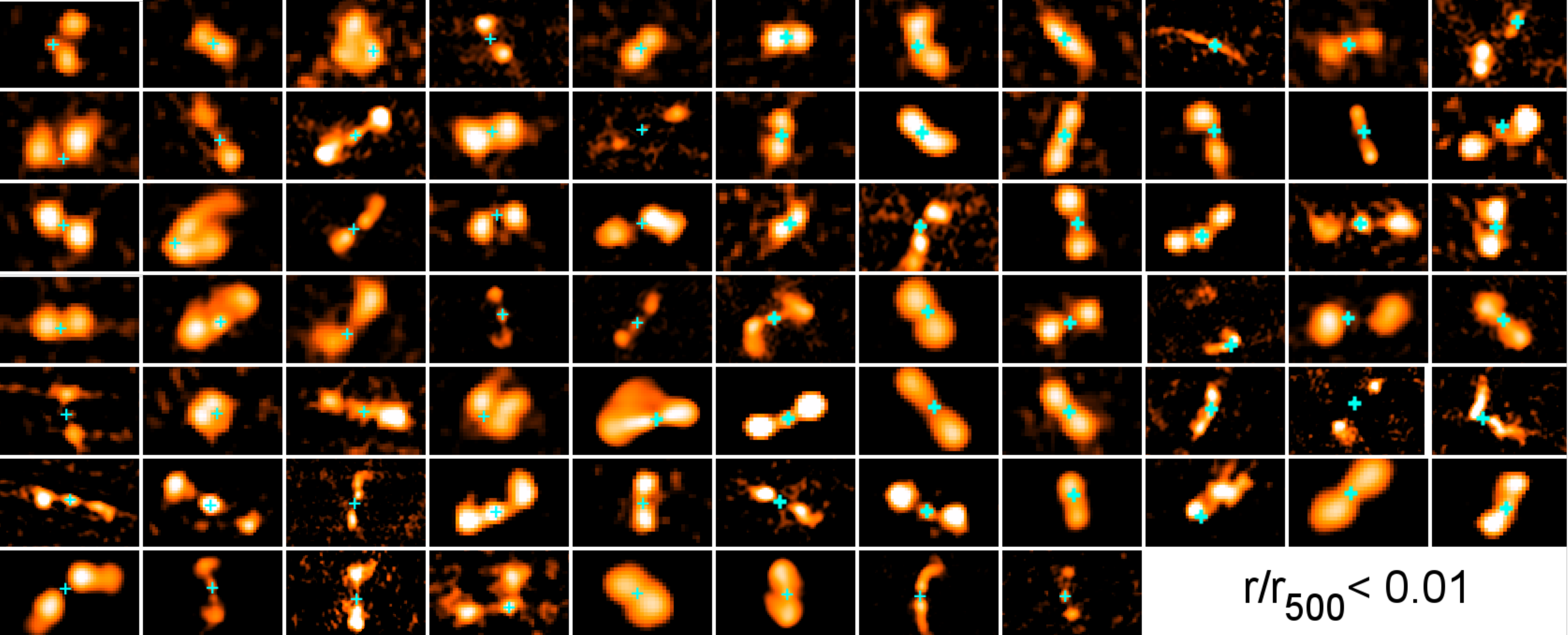} \\
\vspace{2mm}
\includegraphics[width=0.95\textwidth]{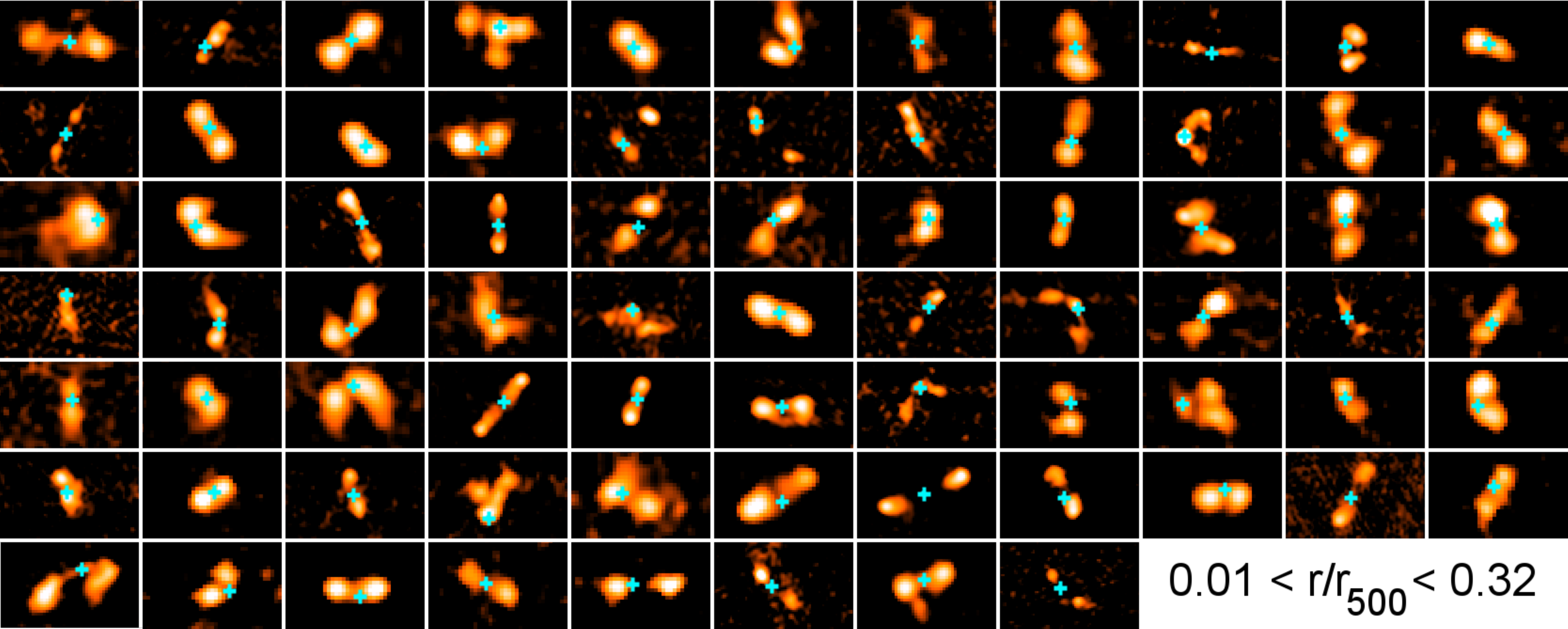} \\
\vspace{2mm}
\includegraphics[width=0.95\textwidth]{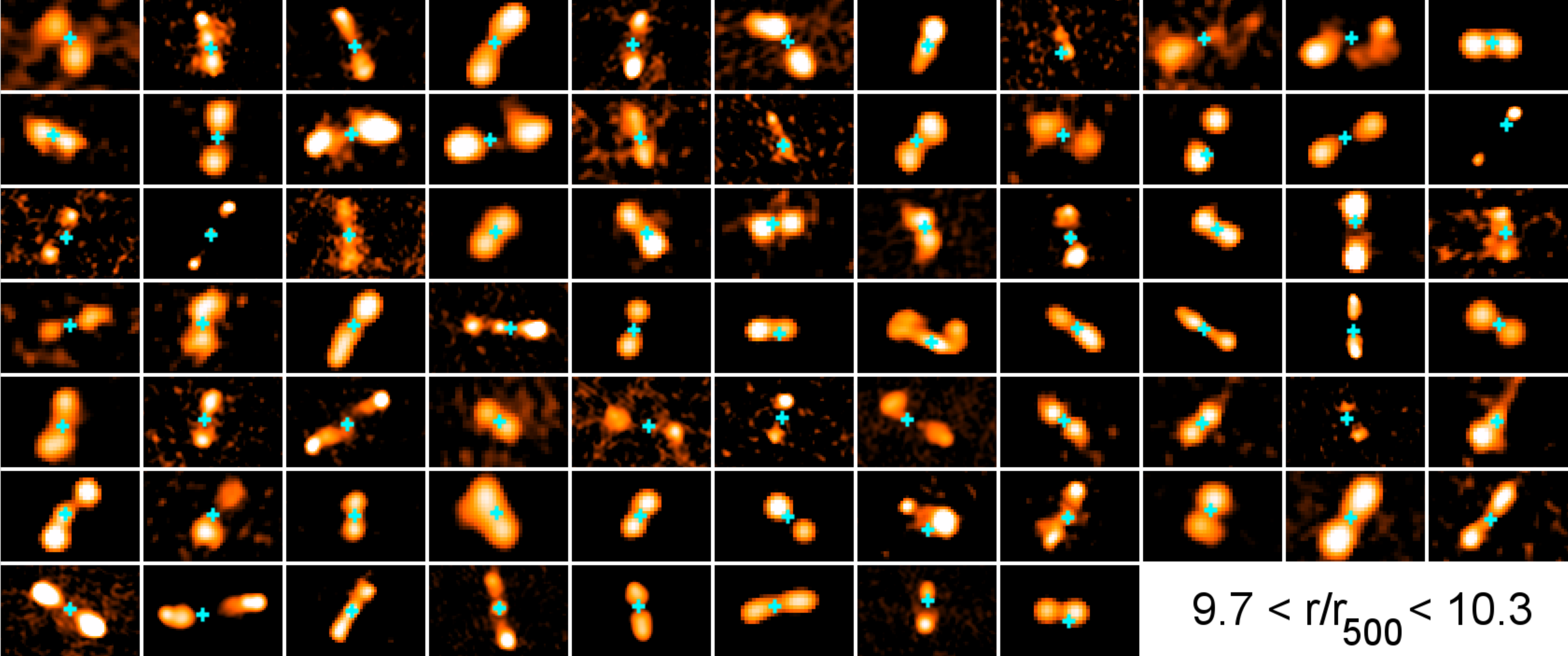}
\caption{Mosaics showing a sample of sources from each region. The host galaxy position is marked.
\emph{From top to bottom:} BCGs, \emph{inner region} sources, \emph{outer region} sources. There are
544, 366, and 2813 sources total in each region, respectively.}
\label{fig:mosaic}
\end{figure*}

We calculate the running median of the excess bending angle, $\Delta\theta$, as a function of
normalized separation from the nearest WH15 cluster, as shown in Figure~\ref{fig:excess_vs_r}. The
shaded region represents the running interquartile range. There are three distinct regions in the
distribution of bending angles as a function of separation, which we separate prior to smoothing and
calculating the trends in all figures that are plotted against separation. Beyond $1.5~r_{500}$, we
do not observe a significant trend between bending and separation, and measure a median
$\Delta\theta\sim 0\degr$ by the definition given in Equation~\ref{eq:correction}; we call this the
\emph{outer region} of the cluster. Sources that are not matched to a WH15 cluster also have median
$\Delta\theta$ consistent with $0\degr$. In Section~\ref{sec:sample_dist}, we identified the sources
within $0.01~r_{500}$ as BCGs, so in all further analysis we combine these sources into a single
data point at $r=0.01~r_{500}$. The rest of the sources lie in what we call the \emph{inner region}
of the cluster, between 0.01 and $1.5~r_{500}$. This also corresponds to approximately the virial
region of the cluster. Our data set contains 544 BCGs, 366 sources in the \emph{inner region}, and
2813 sources in the \emph{outer region}.

We find a median excess bending angle of about $8.0\degr$ for BCGs. The Anderson-Darling (AD) test
is a statistical test used to test if a data sample comes from a specific distribution; we use a
2-sample AD test to rule out that BCGs and sources in the \emph{outer region} are drawn from the
same bending population at the $3.7\sigma$ level. Sources in the \emph{inner region} have a median
excess bending angle of $13\degr$. Between 0.01 and $1.5~r_{500}$, the sources become more bent as
they get closer to the center of the cluster, ranging from $\Delta\theta\sim 0\degr$ at
$r=1.5~r_{500}$ up to a median $\Delta\theta\sim 20\degr$ at $r=0.05~r_{500}$. We use a Spearman's
$\rho$ test to determine that the correlation between separation and excess bending in the
\emph{inner region} is significant at the $4.5\sigma$ level.

Figure~\ref{fig:mosaic} shows a sample of FIRST images of sources from each of the three regions.
(Links to the radio/IR overlays on the RGZ forum\footnote{\url{radiotalk.galaxyzoo.org}} for all
sources in our sample are available in the supplemental data files; see the Appendix for details.)
Visual inspection suggests that BCGs and sources in the \emph{inner region} have similar bending
properties to each other, but are different than sources in the \emph{outer region}. While there are
a few approximately straight BCGs and \emph{inner region} sources, there are a large number of
significantly bent radio galaxies in these regions. By comparison, there are a small number of
visibly bent galaxies in the \emph{outer region}, but most are close to straight.

\subsection{Excess bending angle versus cluster mass}

\begin{figure}
\centering
\includegraphics[width=0.49\textwidth]{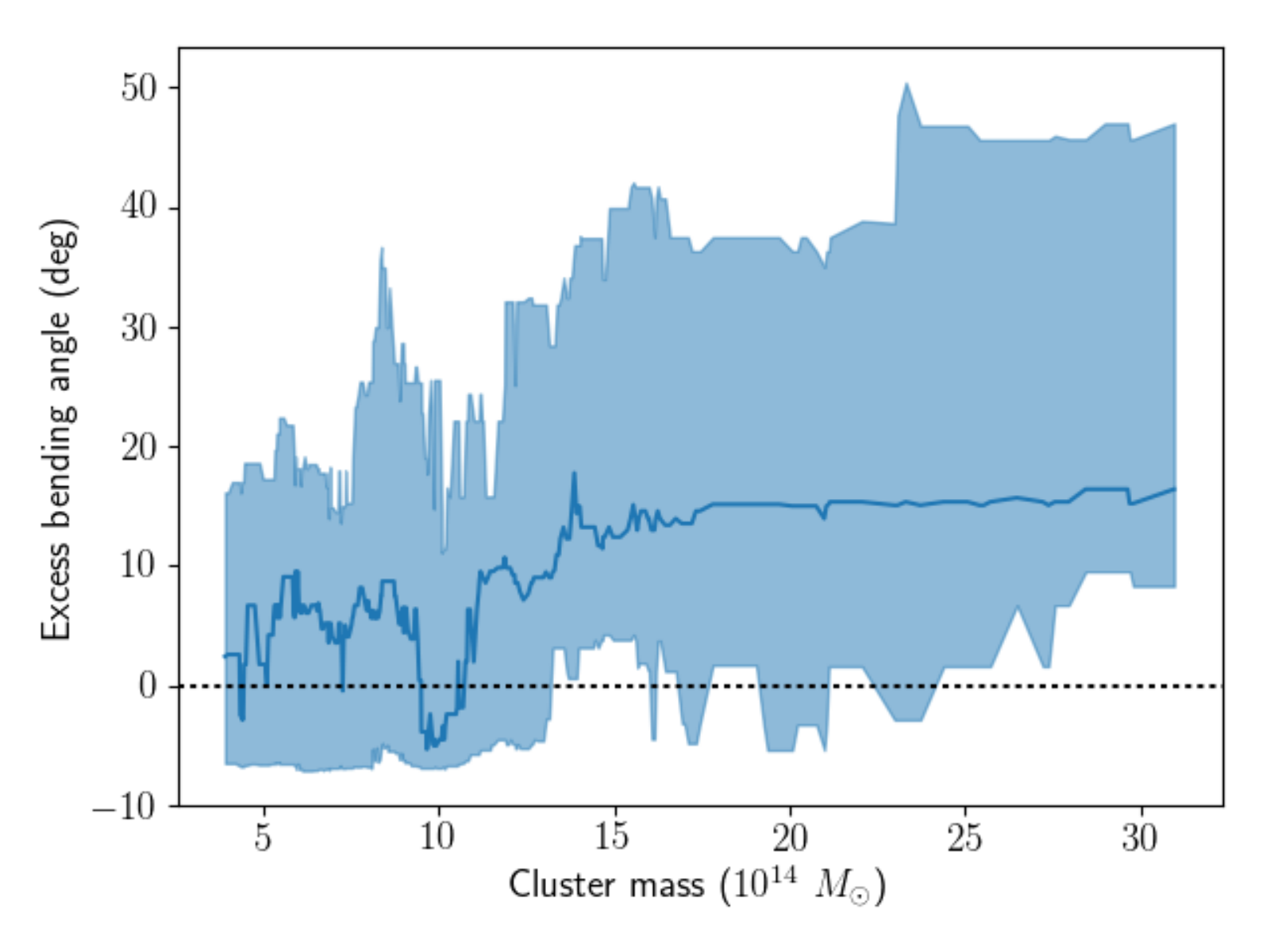}
\includegraphics[width=0.49\textwidth]{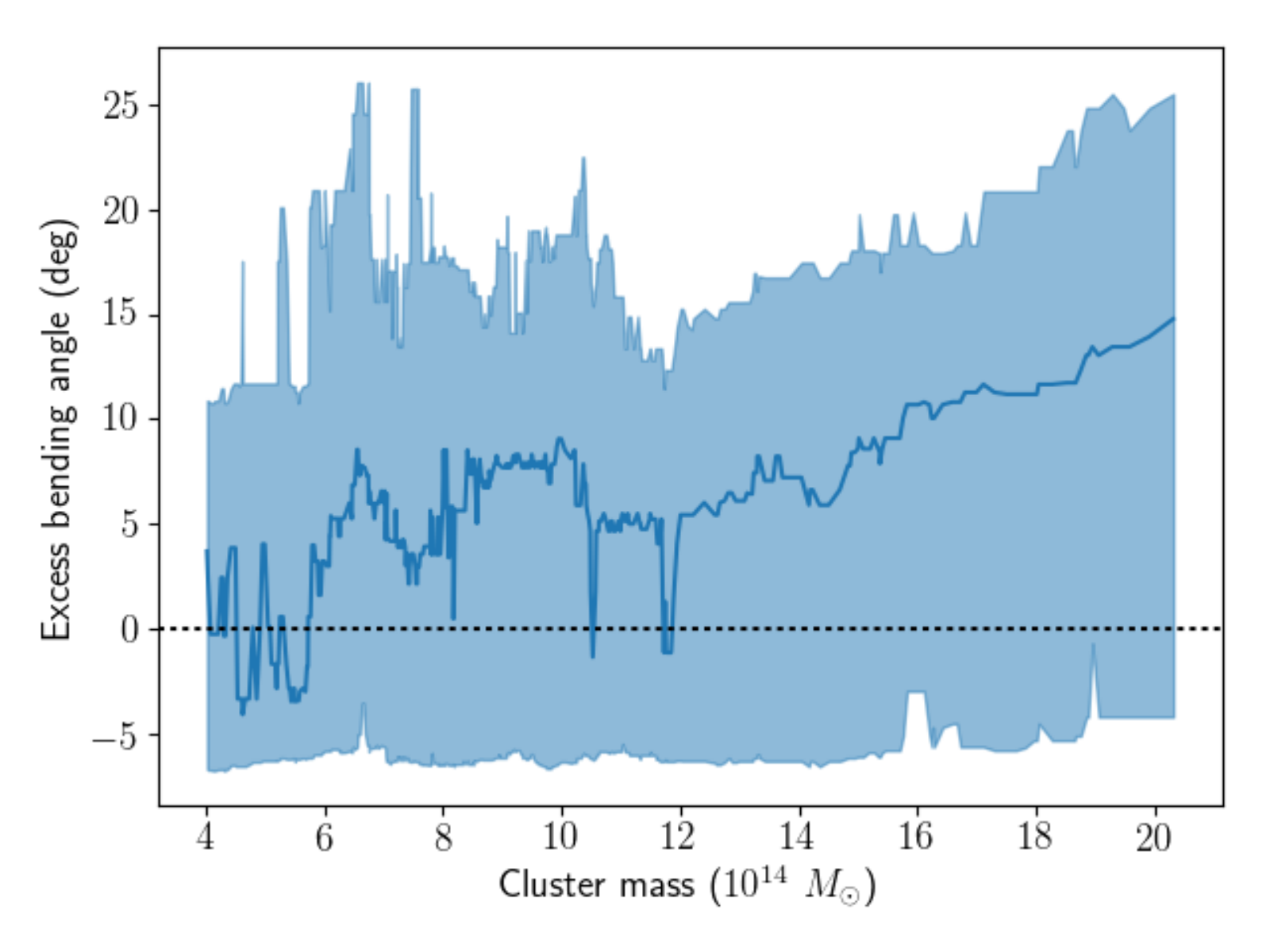}
\caption{Running median of excess bending angle as a function of cluster mass, $M_{500}$. The shaded
region represents the range between the running first and third quartiles. Sources below the dotted
line are consistent with straight. \emph{(Top)} Sources in the \emph{inner region} of clusters
($0.01<r/r_{500}<1.5$). \emph{(Bottom)} BCGs ($r/r_{500}<0.01$).}
\label{fig:excess_vs_m}
\end{figure}

We also expect the mass of the parent cluster to affect the bending of the radio source, since the
orbital velocities of galaxies will be higher in more massive clusters. The velocity dispersion of
galaxies increases with the square root of the mass of the cluster, so ram pressure will be higher
even at a fixed ICM density \citep{becker07,mguda15}. We perform a linear fit between $M_{500}$ and
$r_{500}$ in loglog space for the clusters in our sample and find that $\log_{10} M_{500} \sim
(2.95\pm 0.03) \log_{10} r_{500}$, i.e. the cluster mass is directly proportional to its volume, so
we argue that the ICM density is independent of cluster mass to first order.

The influence of cluster mass on the 366 sources in the \emph{inner region} of the cluster is shown
in the top plot of Figure~\ref{fig:excess_vs_m}, because that is where we find the cluster has a
statistically detectable effect on the bending. We use a Spearman's $\rho$ test and find that the
correlation between cluster mass and excess bending angle is significant at the $4.2\sigma$ level.
We observe an increase in the median $\Delta\theta$ of $10\degr$ between cluster masses of 5 and
$20\times 10^{14}~M_\sun$.

We also look at the cluster mass relationship for the 544 BCG sources separately, shown in the
bottom plot of Figure~\ref{fig:excess_vs_m}. There is a suggestive correlation between cluster mass
and excess bending angle for the BCGs as well. We use a Spearman's $\rho$ test to determine that the
correlation between cluster mass and excess bending angle is significant at the $2.5\sigma$ level
for the BCG sample, and we observe the same increase in median $\Delta\theta$ of $10\degr$ across
the same range of cluster masses.

\subsection{Excess bending angle versus pressure} \label{sec:results_pressure}

The bending angle of radio emission depends on both the mass of the galaxy cluster and the position
of the radio source within the cluster, as shown above. We assume that the observed bending is
induced by the ram pressure exerted on the galaxy by its motion through the ICM, $P_{ram} =
\rho_{icm} v_{gal}^2$ \citep{begelman79,jones79}, although this argument is equivalent if it is an
ICM wind flowing past the galaxy \citep{burns02}. In either case, $\rho_{icm}$ is the ICM density
and $v_{gal}$ is the relative velocity between the galaxy and the ICM. Here we use $P_{icm}$, the
ICM gas pressure, as a proxy for $P_{ram}$, under the assumption that the clusters are virialized to
first order. In that case, the velocity of the galaxies and the velocity of the ICM gas particles
both scale with the temperature of the ICM. We justify this in more detail in
Section~\ref{sec:bending_disc}.

\cite{arnaud10} derived a ``universal galaxy cluster pressure profile'' using a sample of local
clusters observed with \emph{XMM-Newton}, to which they fit a generalized NFW profile. They derive
\begin{multline} \label{eq:pressure}
P_{icm}(x) = 1.65\times 10^{-3}~h(z)^{8/3}~\mathbb{P}(x)~\textrm{h}_{70}^2~\textrm{keV~cm}^{-3} \\
\times \left(\frac{M_{500}}{3\times 10^{14}~M_\sun} \textrm{h}_{70}
\right)^{2/3+\alpha_P+\alpha'_P(x)}
\end{multline}
as the best universal pressure relation, where $x=r/r_{500}$, $\mathbb{P}(x)$ is the generalized NFW
model from \cite{nagai07}, and $\alpha_P$ and $\alpha'_P(x)$ are derived from the slope of the
$M_{500}-Y_X$ relation.

\begin{figure}
\centering
\includegraphics[width=0.49\textwidth]{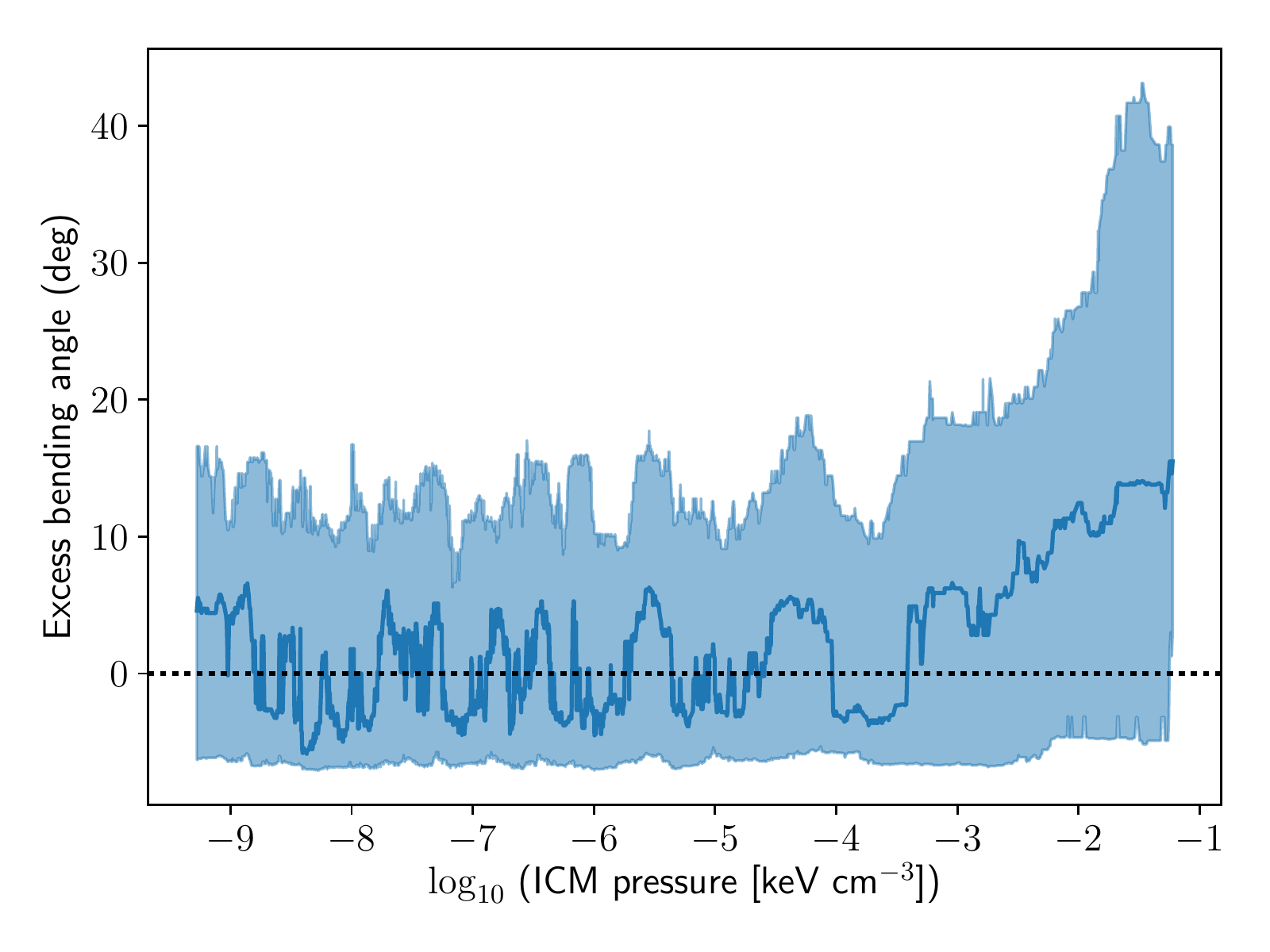}
\caption{Running median of excess bending angle as a function of inferred ICM gas pressure,
$P_{icm}$. The shaded region represents the range between the running first and third quartiles.
BCGs have been excluded. Sources below the dotted line are consistent with straight.}
\label{fig:excess_vs_p}
\end{figure}

We apply Equation~\ref{eq:pressure} to our sample to get a measure of the local ICM gas pressure.
Figure~\ref{fig:excess_vs_p} shows the distribution of excess bending angle as a function of
pressure, BCGs excluded. As shown in the previous plots of separation and mass, low pressure
($<5\times 10^{-4}~\textrm{keV~cm}^{-3}$) environments (i.e. at large separations [low density]
and/or in low mass clusters [low velocity]) do not induce a statistical increase in bending. Above
$5\times 10^{-4}~\textrm{keV~cm}^{-3}$, increasing the pressure leads to a greater excess bending
angle, up to a median $\Delta\theta\sim 14\degr$ when the pressure reaches
$10^{-2}~\textrm{keV~cm}^{-3}$.

\subsection{Orientation} \label{sec:results_orient}

\begin{figure}
\centering
\includegraphics[width=0.49\textwidth]{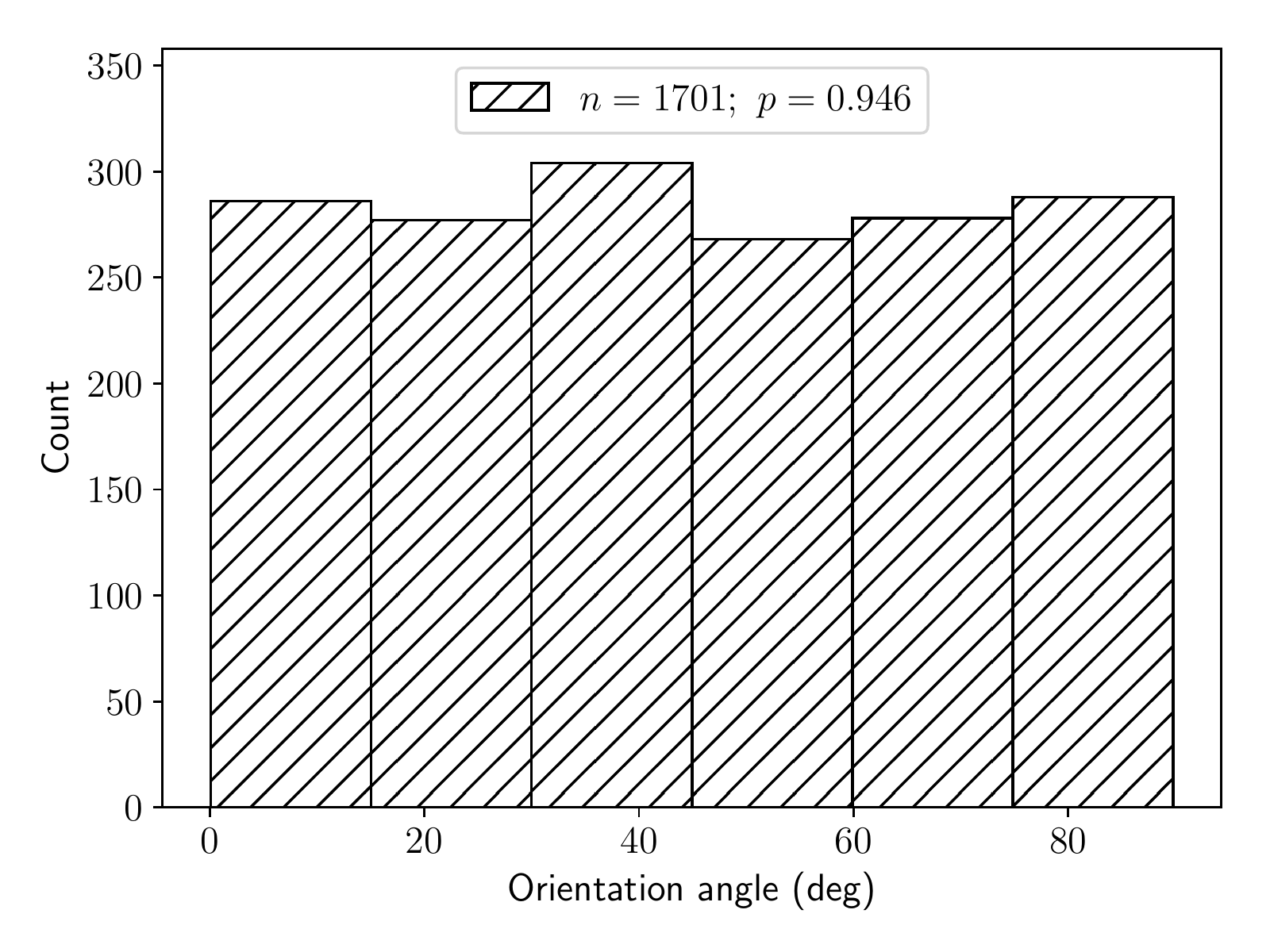}
\includegraphics[width=0.49\textwidth]{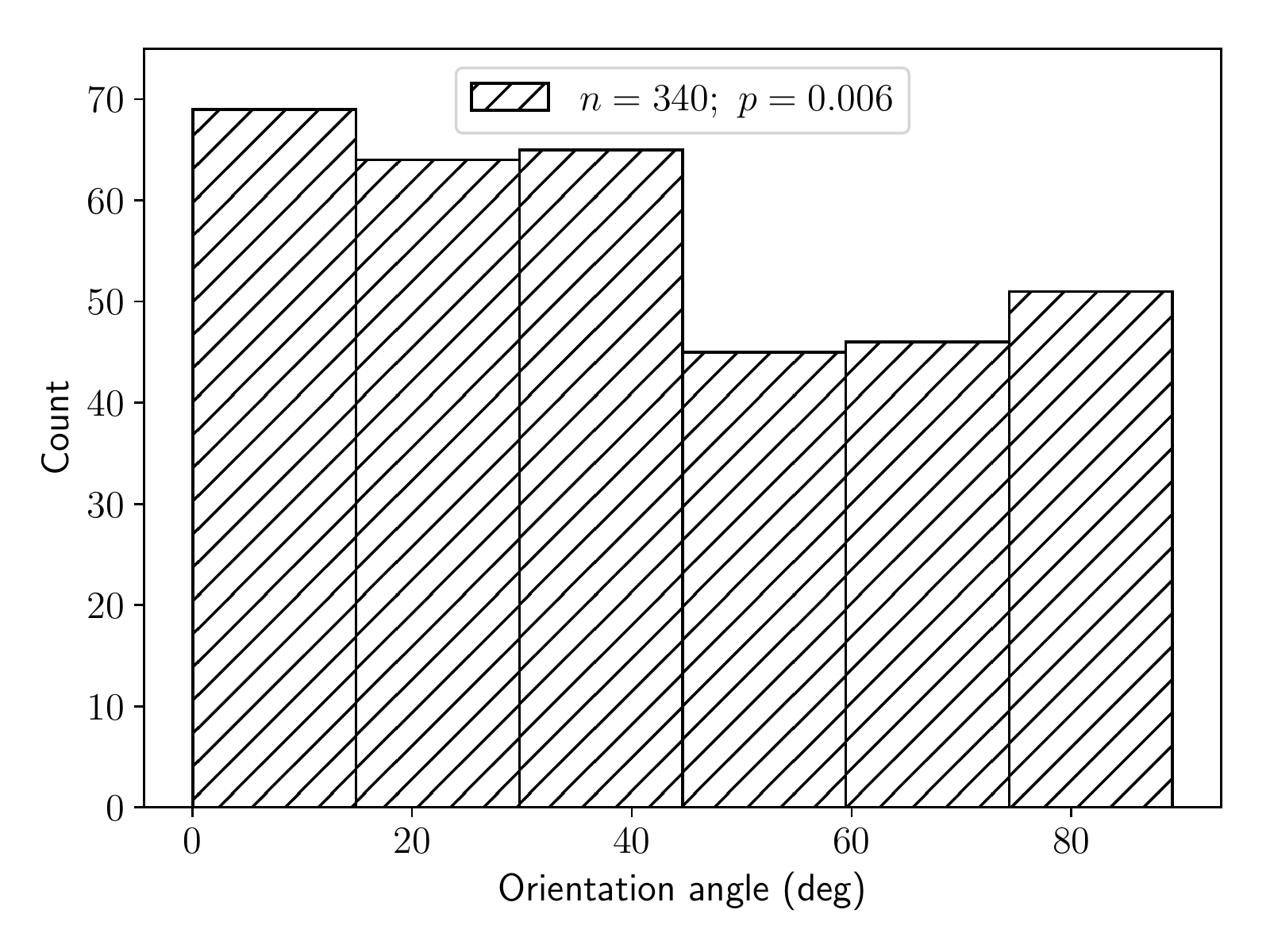}
 \caption{Histograms of orientation angle, folded around $90\degr$, for highly bent and less bent
sources between 0.01 and $10~r_{500}$. Given $p$-values correspond to the null hypothesis that the
distribution of orientations is uniform. \emph{(Top)} Less bent sources are uniformly distributed in
orientation angle. \emph{(Bottom)} The distribution of orientation angles for highly bent sources
peaks near $0\degr$ (radial) and decreases at larger orientation angles.}
\label{fig:orientation}
\end{figure}

We define the orientation angle of a source by taking the difference between the angle bisector of
the opening angle and the radial vector from the center of the cluster. For the purpose of this
analysis, we used the ``folded'' orientation, ignoring whether the source opened towards or away
from the cluster. This means that the orientation angle ranges from $0\degr$ if the source opened
parallel to the radial vector from the cluster (either towards or away), and increases to $90\degr$
if it opened perpendicular to the radial vector.
 
We measure the distributions of orientation angles for two sub-populations: highly bent sources
($\Delta\theta>21\degr$) and less bent sources ($\Delta\theta<21\degr$). We choose $21\degr$ as the
fiducial threshold for significant bending as this captures the 16\% most highly bent sources in our
full matched sample. Both sub-populations are restricted to sources matched between 0.01 and
$10~r_{500}$. We exclude sources beyond $10~r_{500}$ to keep the contamination below a few percent.
To determine whether these samples were randomly distributed in orientation angle, we compared the
distributions to the uniform distribution using a 1-sample AD test. The distributions and
corresponding $p$-values are shown in Figure~\ref{fig:orientation}. The less bent sources are
uniformly distributed in orientation. However, highly bent sources tentatively appear to be
preferentially aligned parallel to the radial vector; we reject that they are uniformly distributed
in orientation at the $2.7\sigma$ level.

\begin{figure}
\centering
\includegraphics[width=0.49\textwidth]{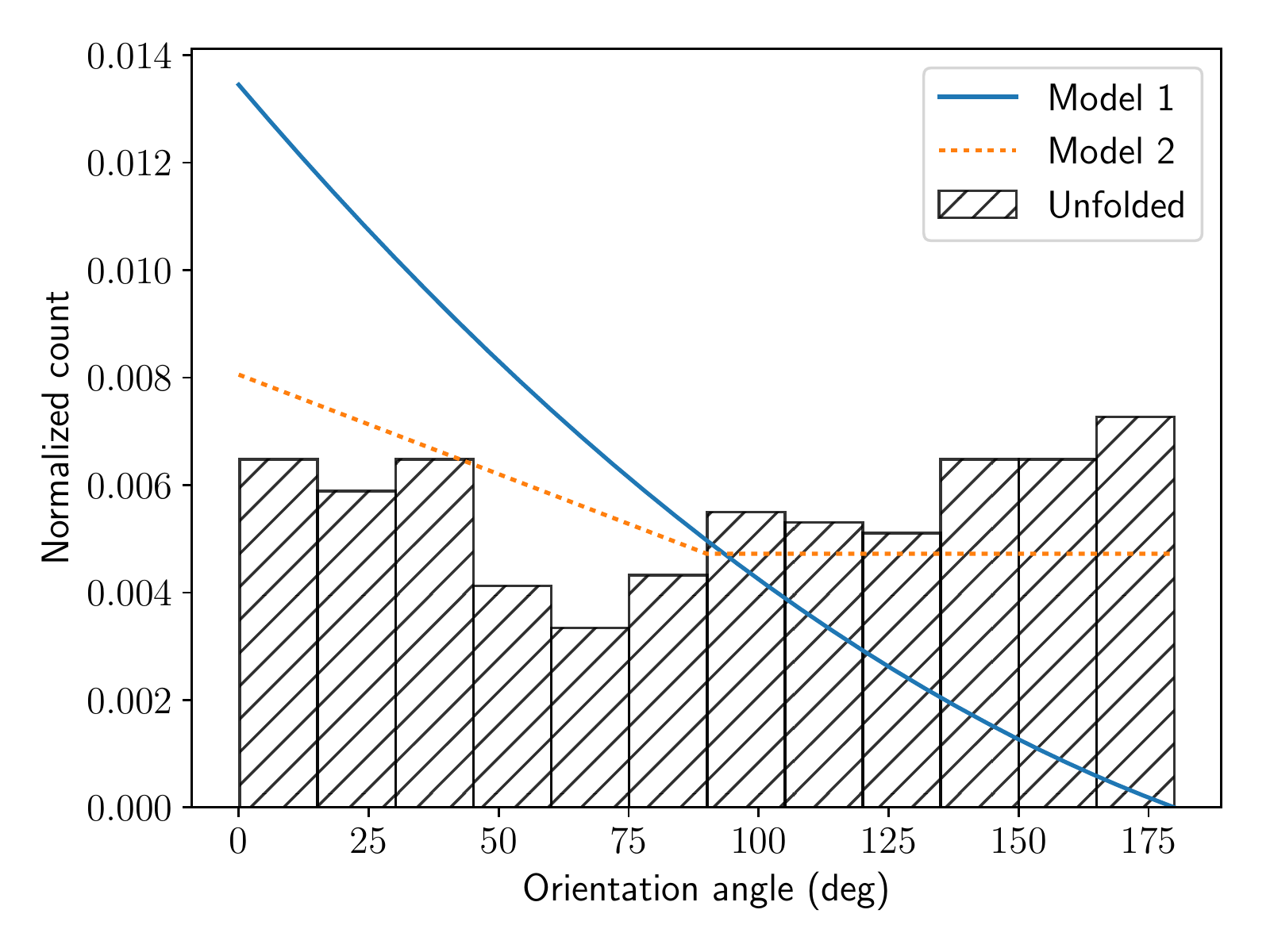}
\caption{Normalized histogram of orientation angle for highly bent sources between 0.01 and
$10~r_{500}$, unfolded. Model~1 (solid) and Model~2 (dotted) are constructed distributions due to
buoyancy described in the text. We rule out the unfolded distribution being drawn these models at
the $4.8\sigma$ and $4.1\sigma$ level, respectively.}
\label{fig:orientation_test}
\end{figure}

To address the question of whether buoyancy may explain the excess of radially-oriented highly bent
sources, we also looked at the ``unfolded'' orientation angles. Observationally, if the excess of
radially-oriented sources was due to buoyancy, then the unfolded distribution would peak near
$0\degr$ and drop off near $180\degr$. Theoretically, pressure gradients such as those that generate
buoyant forces should be less important than ram pressure in the bending of radio sources.

We generate models of the two extremes of the probability density function (PDF) of orientations
that could arise solely due to buoyancy, which are plotted in Figure~\ref{fig:orientation_test}
alongside the unfolded distribution. Both distributions are monotonic and peak at orientations of
$0\degr$, and are constructed so as to recover the observed folded distribution in the bottom plot
of Figure~\ref{fig:orientation}. Model~1 is a quadratic PDF that peaks for sources opening away from
the cluster and decreases to 0 for sources opening towards the cluster; it represents the physical
situation where buoyancy strongly dominates. Model~2 is a piecewise linear PDF that also peaks for
sources opening away from the cluster but plateaus for sources that point inward of tangential; it
represents the physical situation where buoyancy contributes the minimum necessary to reproduce the
folded distribution. We compare the observed unfolded distribution to both models using a 1-sample
AD test, and rule out it having been drawn from Model~1 at the $4.8\sigma$ level and from Model~2 at
the $4.1\sigma$ level. As these models cover the range of buoyancy-driven PDFs, we rule out buoyancy
as driving the preference for radial orientation.

\subsection{Distribution of bent sources}

\begin{figure}
\centering
\includegraphics[width=0.49\textwidth]{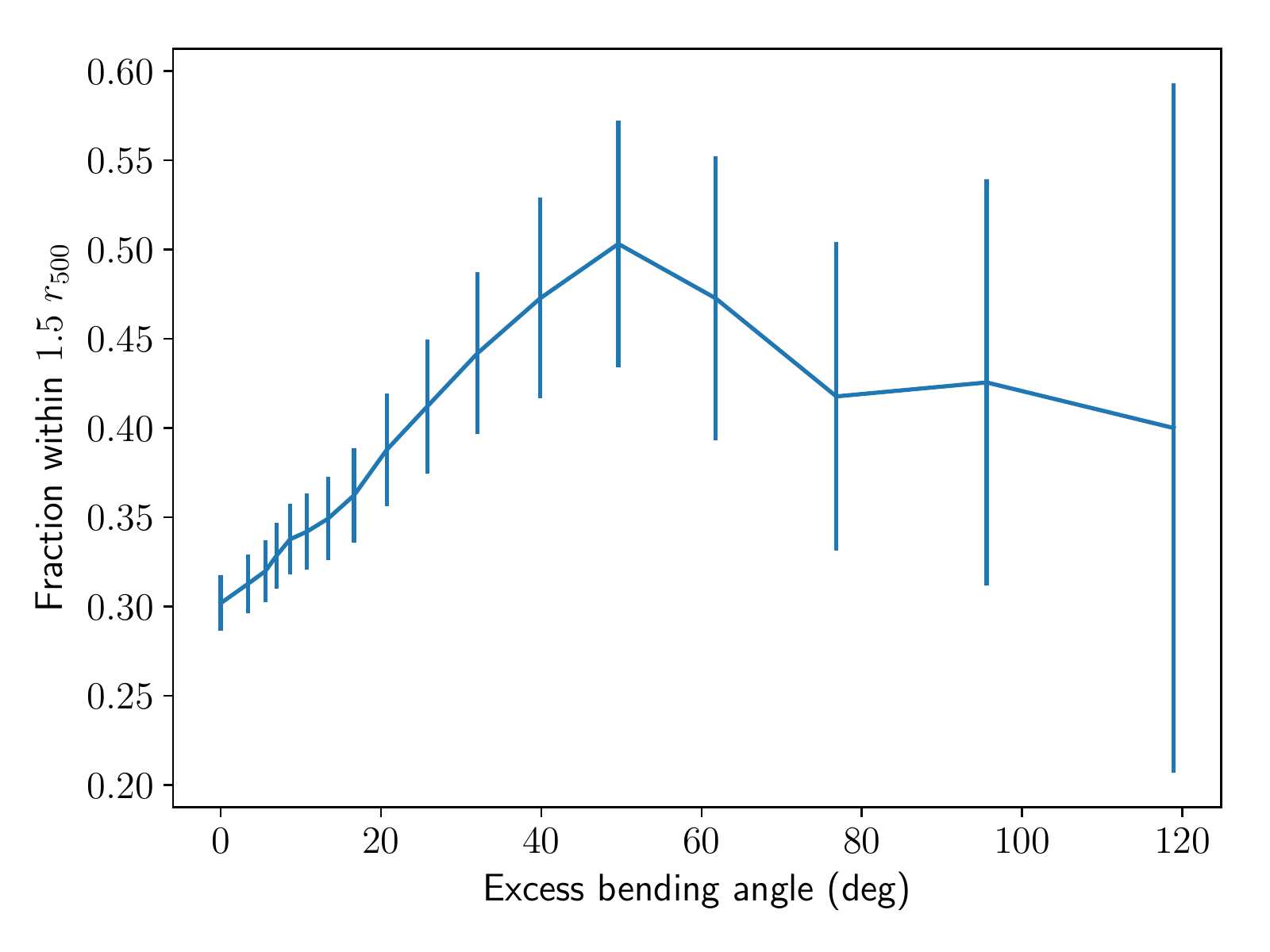}
\caption{Ratio of the number of sources matched to within $1.5~r_{500}$ of a cluster and the number
of all sources in our sample (including sources with no WH15 cluster match within a projected
15~Mpc), for a given minimum excess bending angle. Even at large excess bending angles, the majority
of sources are in the \emph{outer region} of the nearest cluster or beyond.}
\label{fig:bending_fraction}
\end{figure}

While proximity to a cluster strongly affects the bending angle of a radio source, this is only a
statistical excess. Other properties of the source also affect how the jets are bent; for example,
the Mach number of the jets, which is not directly measurable, should negatively correlate with the
amount of bending \citep{jones17}. In order to assess how much of the bending can be explained using
proximity alone, we show in Figure~\ref{fig:bending_fraction} the fraction of sources in our sample
above a given excess bending angle that are within $1.5~r_{500}$ of the nearest cluster. Here we
also include in the total sources not matched to a WH15 cluster at all. While the WH15 catalog goes
out to $z=0.8$, the completeness of SDSS (and thus our unmatched sample) begins to fall off beyond
$z=0.6$ \citep{beck16}; to avoid biasing our results due to this, we restrict the analysis in this
section to sources with $z<0.6$.

As shown in Figure~\ref{fig:bending_fraction}, $(27\pm 1)\%$ of all sources are within
$1.5~r_{500}$. If proximity to a cluster center were the dominant source of bending, we might expect
almost all highly bent sources to be found within 1.5 r500. However, for sources with
$\Delta\theta>21\degr$, we find only $(40\pm 5)\%$ within this range. Even at the largest bending
angles, most sources are beyond $1.5~r_{500}$: the distribution peaks at $\Delta\theta>50\degr$, for
which still only $(50\pm 10)\%$ of sources are within $1.5~r_{500}$. These values are similar to
those of \cite{blanton01}. Simply due to the large number of radio sources in our sample that are
not in the interior of clusters, any particular bent source is as likely as not to be far from a
cluster. While a highly bent radio galaxy is suggestive of a nearby cluster, its membership in a
cluster requires follow-up observations of the environment.

\subsection{Environment of \emph{outer region} bent sources}

Some mechanism must still be bending sources, even away from clusters. To take an initial look at
this issue, we examined the optical environments of two subsets of radio galaxies at a median
separation of $9~r_{500}$ from the nearest cluster. The first sample consists of 150 highly bent
sources (median $\Delta\theta\sim 46\degr$); the second consists of 150 less bent sources (median
$\Delta\theta\sim -7\degr$). For each of these radio galaxies, we search SDSS for nearby galaxies
within a redshift range of $\pm0.04(1+z)$ and within a projected radius of $0.33\degr$, and stack
the resulting counts as a function of projected physical separation.

\begin{figure}
\centering
\includegraphics[width=0.49\textwidth]{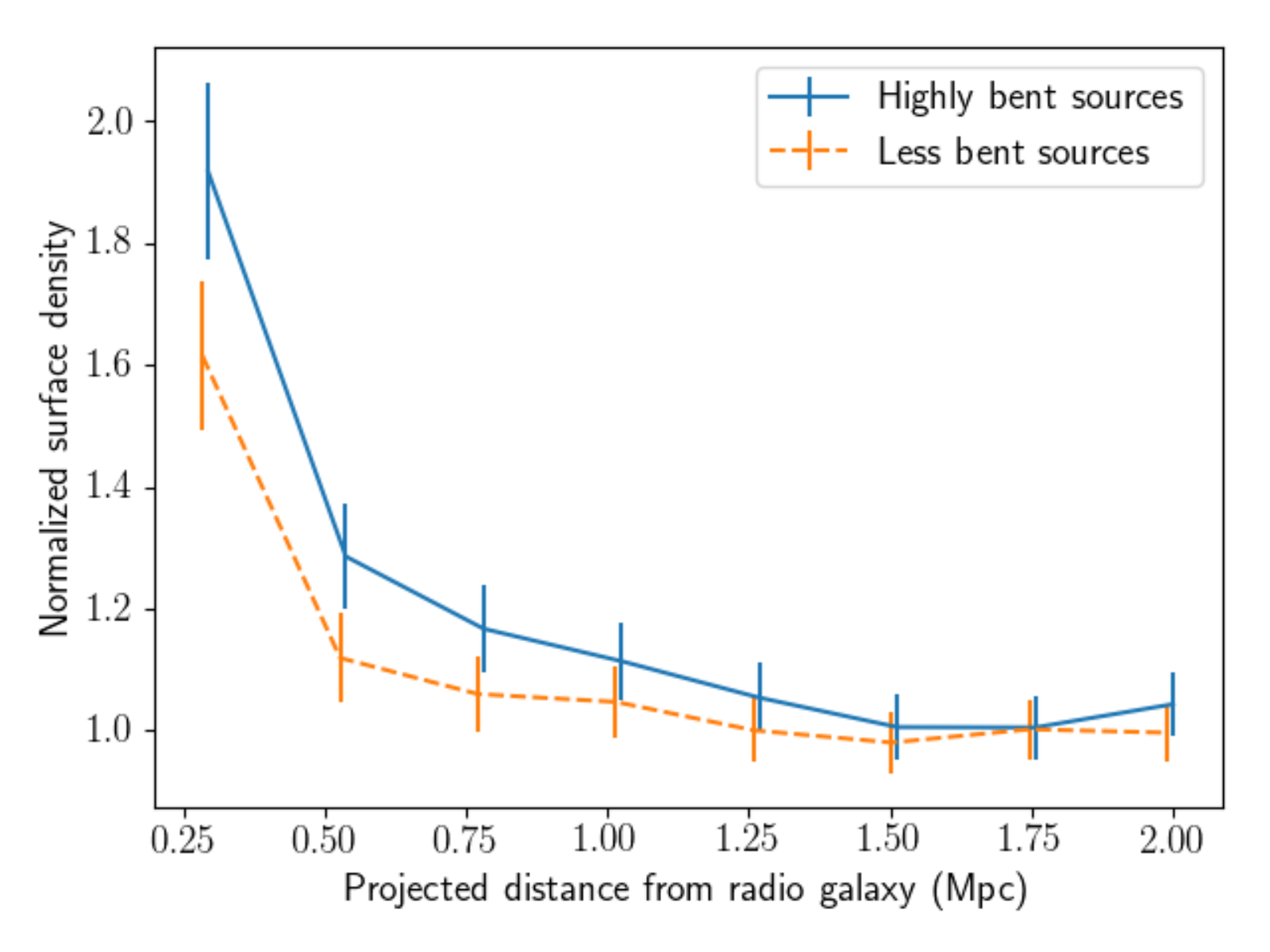}
\caption{Surface density of optically-detected galaxies around radio sources outside of clusters, i.e. sources $\sim 9~r_{500}$ from the nearest cluster, normalized by the background surface density. The background surface density of galaxies is 6.7~Mpc$^{-2}$ around highly bent sources and 7.4~Mpc$^{-2}$ around less bent sources. Highly bent sources are shown with the solid line, and less bent sources are shown with the dashed line. The error bars for less bent sources are offset for legibility.}
\label{fig:distant_env}
\end{figure}

Since we are no longer near the center of a cluster, we measure the separation between a radio
galaxy and its optical neighbors in Mpc instead of scaling by $r_{500}$. We find that the surface
density of galaxies neighboring the radio galaxy plateaus beyond a projected 2~Mpc from the radio
galaxy, so we define the background surface density as the median surface density in an annulus
between 2 and 3~Mpc. This is 6.7~Mpc$^{-2}$ around our sample of highly bent sources and
7.4~Mpc$^{-2}$ around our sample of less bent sources. Figure~\ref{fig:distant_env} shows the
surface density as a function of separation from the radio source, normalized by the background
surface density.

The density excess rises sharply within 350~kpc of the radio sources. Between 50 and 350~kpc, we estimate an enhancement in volume density by a factor of $4.7\pm 0.6$ for highly bent sources and $2.9\pm 0.5$ for less bent sources compared to the local background density of SDSS galaxies. Assuming an $r^{-2}$ scaling of volume density as a function of distance to the nearest cluster center as inferred from Figure~\ref{fig:density}, these values suggest that the radio galaxies are in local densities of approximately $28 \rho_{crit}$ and $18 \rho_{crit}$ for highly bent and less bent sources, respectively. Such concentrations of galaxies are not included in the WH15 catalog because they are not dense enough to exceed their richness threshold of $R_{L*,200}\geq 12$.

\subsection{Radio tail asymmetry} \label{sec:results_asym}

We look at the effect of cluster environment on the asymmetry of the radio jets. Given a radial density gradient in the ICM, the inward-pointing jet of a radially-aligned radio source will be in a higher density environment and should be suppressed compared to the outward-pointing jet. While a given cluster can have a clumpy ICM, this effect should be detectable statistically.

This analysis complements that of \cite{rodman19}, who performed a detailed examination of the morphology and the galactic environment for each source in their sample, although their sample only contained 23 sources. Here, without examining the sources individually, we measure the length of each tail using the distance between the host galaxy and the most distant point on the lowest radio contour that does not contain the host. (This is the same method we used to calibrate the bending angle uncertainty in Section~\ref{sec:correction}.)

We separate our sample into two sub-populations: a) the 120 \emph{inner region} sources that are aligned radially with the cluster, defined as ones where one tail is within $45\degr$ of the direction towards the cluster center, and the other is within $45\degr$ of the direction away from the cluster center; and b) the 125 \emph{inner region} sources that are aligned tangentially to the cluster, defined as ones where both tails are within $45\degr$ of tangential to the direction of the cluster center. We define the asymmetry of the source as the ratio of the length of the inward-pointing tail to the length of the outward-pointing tail (even tangentially-aligned sources have one tail that is more inward-pointing than the other).

For radially-aligned sources, we observe that the inward-pointing tail is shorter than the outward-pointing tail, and that the amount of asymmetry increases for sources closer to the center of the cluster. We fit a line to the trend and calculate that it has a slope of $0.18\pm 0.09~r_{500}^{-1}$, and a Spearman's $\rho$ test determines the correlation between separation and asymmetry is significant at the $2.1\sigma$ level. By comparison, we do not observe a statistically significant correlation for tangentially-aligned sources; Spearman's $\rho$ test results in only a $1.5\sigma$ result.

We look at the jet asymmetry using pressure as well. The asymmetry should depend on the pressure gradient across the source, which we approximate as the difference in pressure at the ends of the two tails divided by the radial component of the source size. Using the same sample of radially-aligned sources, the correlation between pressure gradient and asymmetry is lower than when looking at separation only: only $1.7\sigma$. Our simple pressure model does not account for the asymmetry as well as just using the position within the cluster.

\subsection{Summary of results} \label{sec:summary}

\begin{enumerate}
\item Both extended radio galaxies and optically-selected galaxies follow approximately isothermal distributions in their separations from cluster centers centers, but the power law index for radio galaxies ($\alpha = -1.10\pm 0.03$) is slightly but significantly steeper than the one for optically-selected ones ($\alpha = -0.94\pm 0.02$).
\item BCGs exhibit extended radio emission at least $2.5\pm 0.3$ times more often than the general population of cluster galaxies.
\item Median excess bending angle ($\Delta\theta$), defined in Equation~\ref{eq:correction}, increases closer to the center of clusters. Sources are statistically consistent with straight beyond $1.5~r_{500}$, but increase to a median excess bending angle of $\Delta\theta\sim 24\degr$ at $0.01~r_{500}$.
\item BCGs have median $\Delta\theta\sim 8\degr$. This makes them more bent than non-cluster sources, but less bent than the innermost non-BCG sources.
\item For BCGs and sources in the \emph{inner region} of clusters, $\Delta\theta$ correlates with cluster mass. The median $\Delta\theta$ for BCGs increases by $\sim 10\degr$ between cluster masses of 5 and $20\times 10^{14}~M_\sun$; for \emph{inner region} sources, it also increases by $\sim 10\degr$ over the same range of masses.
\item For non-BCG sources, $\Delta\theta$ correlates with local ICM gas pressure (a proxy for ram pressure) above $\sim 5\times 10^{-4}~\textrm{keV~cm}^{-3}$. Below this pressure, sources are statistically consistent with straight.
\item Highly bent sources ($\Delta\theta>21\degr$) preferentially have opening angles oriented parallel to the radial vector to the cluster center, but there is no particular preference between opening towards or away. Less bent sources ($\Delta\theta<21\degr$) do not exhibit a preferential orientation.
\item While BCGs and sources in the \emph{inner region} of clusters are more likely to have highly bent angles than low ones, for any given bending angle, there are more sources outside of clusters than inside.
\item There is an excess of optically-selected neighbor galaxies within 350~kpc around radio galaxies far from clusters (median cluster distance $\sim 9~r_{500}$), and the excess is larger around highly bent sources: $4.7\pm 0.6$ and $2.9\pm 0.5$ times the background volume density for highly bent and less bent sources, respectively. This suggests that the radio galaxies are in local densities of approximately $28\rho_{crit}$ and $18\rho_{crit}$, respectively.
\item Radio tail length asymmetry exhibits a low significance ($2.1\sigma$) dependence on separation from the cluster center. Among \emph{inner region} sources that are aligned with their tails parallel to the radial vector, the inward-pointing tail is shorter than the outward-pointing tail, and the difference increases for sources closer to the center of the cluster. Asymmetry exhibits a similar, but even less significant ($1.7\sigma$), dependence on the pressure gradient across the source.
\end{enumerate}

\section{Discussion} \label{sec:discussion}

\subsection{Distribution of radio galaxies}

In generating our sample of radio galaxies, we mapped out their projected spatial distribution around clusters and found that they appear to be distributed according to a power law similar to the distribution of optically-selected galaxies in clusters. This implies that we are not catching them at special times (e.g. if they were triggered when first encountering the outskirts of clusters). However, the power law distribution is steeper for radio galaxies (although we do not account for differences caused by mass segregation and defer this until future analysis). Previous work investigating the angular two-point correlation function, $\omega(\theta)$, of radio sources on the sky and comparing this to $\omega(\theta)$ of a matched sample of radio-quiet galaxies has also found that radio sources exhibit greater clustering than do radio-quiet galaxies \citep{wake08}.

\cite{pimbblet13} investigated if they could detect a preferential distribution of active galaxies (AGN) in clusters. They did not find a difference between AGN and the general population. This suggests that while AGN activity in general is not preferentially triggered by interactions with other galaxies (at least not on a timescale before the host galaxy has moved away from the interaction site), the AGN is more likely to be radio-loud in the proximity of other galaxies. \cite{pimbblet13} cross-referenced FIRST as well and found no relationship between radio luminosity and position within the cluster, but they noted that their sample size of 28 was too small to draw a firm conclusion.

The connection between radio-loud AGN and cluster environment is also supported by \cite{chiaberge15}, who investigated the merger history of radio-loud and radio-quiet AGN. They found that radio-loud AGN are almost always the product of merging systems, whereas only a fraction of radio-quiet AGN experienced a recent merger. From a clustering perspective, galaxy interactions will happen more often when the density of galaxies is higher, thus connecting our result that radio galaxies are preferentially found near other galaxies to the result of \cite{chiaberge15}.

\subsection{Radio galaxy bending} \label{sec:bending_disc}

The primary goal of this work was to investigate the influence of the cluster on the bending of radio galaxies. Bending angle of a radio source is not a well-defined physical quantity; the definition used in this paper is a somewhat arbitrary approximation of a source's physical radius of curvature, and even that breaks down for highly disturbed sources. While it is an insufficient parameter to compare against physical models of individual sources, it is sufficient for determining statistical trends in morphology.

We observe two physical properties of the cluster that correlate statistically with the bending of the source: the position of the galaxy within the cluster and the mass of the cluster. Using the model of ram pressure bending first invoked by \cite{miley72} and quantified by \cite{begelman79} and \cite{jones79}, we interpret these correlations as physically corresponding to the two components of ram pressure: the ICM density and the relative velocity between the galaxy and ICM, respectively.

Within the \emph{inner region} of the clusters ($0.01<r/r_{500}<1.5$), which is approximately equal to the virial region (excluding the BCG), we find that sources closer to the center have a larger median excess bending angle ($\Delta\theta$, defined in Equation~\ref{eq:correction}) than sources farther away. Beyond $1.5~r_{500}$, the measured bending angle is comparable to the uncertainty in our bending measurement. Close to the center of the cluster, the density of the ICM quickly rises \citep{pratt02,newman13}, which leads to the increase in $\Delta\theta$ of sources in the denser surroundings.

Also restricted to the \emph{inner region}, sources in higher mass clusters are more bent than sources in lower mass clusters. The velocity dispersion of the galaxies in a cluster is proportional to the square root of the cluster mass via the gas temperature \citep{becker07,mguda15}, so $M_{500}$ serves as a proxy for the relative velocity between the source and the ICM. We find that $M_{500} \propto r_{500}^{2.9}$ for the clusters in our sample, so we argue that the ICM density is constant to first order as a function of cluster mass and does not play a significant role in our result: among sources that are already in a sufficiently dense medium, the ones with statistically higher velocities are statistically more bent.

Together, these two trends correspond to a statistical increase in ram pressure causing statistically greater bending. In Section~\ref{sec:results_pressure}, we used the local gas pressure as another proxy for this ram pressure. The effect of ram pressure can be quantified by
\begin{equation} \label{eq:force_balance}
\frac{\rho_j v_j^2}{R} = \frac{\rho_{icm} v_{gal}^2}{h},
\end{equation}
where $\rho_j$ and $\rho_{icm}$ are the jet and ICM densities, respectively; $v_j$ is the jet velocity; $v_{gal}$ is the relative velocity between host galaxy and the ICM; $R$ is the radius of curvature of the jet axis; and $h$ is the jet radius \citep{begelman79,jones79}.

If the cluster is approximately virialized, then both the gas thermal velocity ($v_{icm}$) and the galaxy velocity scale similarly with the ICM temperature, $T_{icm} \propto v_{icm}^2 \propto v_{gal}^2$ \citep[e.g.][]{lubin93}, justifying our use of the static pressure $P_{icm} \propto \rho_{icm} T_{icm}$ as a proxy for the ram pressure $\rho_{icm} v_{gal}^2$. As long as the jet radii ($h$) and lengths ($\ell$) are statistically independent of the cluster properties, then the bending angle ($\theta$) will scale with $P_{icm}$ according to
\begin{equation}
\theta \propto \frac{\ell}{R} \propto \frac{\ell}{h} \frac{\rho_{icm} T_{icm}}{\rho_j v_j^2} \propto \left(\frac{\ell}{h \rho_j v_j^2}\right) P_{icm},
\end{equation}
as we have found.

According to the simulations of \cite{marshall18}, AGN activity is triggered when the ram pressure is at least a factor of 2 greater than $P_{icm}$. From this we estimate that the minimum $P_{ram}$ required to induce significant bending is at least of order magnitude $10^{-3}~\textrm{keV~cm}^{-3}$. We compare this to \cite{mguda15}, who used characteristic jet values from \cite{freeland11} to estimate that ram pressures above $6\times 10^{-3}~\textrm{keV~cm}^{-3}$ (in our units) were required to induce bending, which is consistent with our result. Since we can statistically detect even very small amounts of bending, we expect our threshold to be lower.

From simulations, bending can also be parameterized in terms of the ratio of Mach numbers of the jet and the ICM shock \citep{jones17}. For our purposes, the Mach numbers of ICM flows and galaxy motions are statistically independent of environment, again assuming that clusters are virialized to first order. In addition, there may be non-environmental properties that give rise to bent radio jets, such as precession of the central black hole \citep{falceta-goncalves10}.

\subsection{BCGs}

The BCGs in our sample are only moderately bent. While WH15 defines the center of the cluster to be the location of the BCG, which would suggest that they are stationary, in general this is not physically accurate. Rather, a large fraction of clusters contain BCGs that are offset from the center of mass of the cluster and have a small peculiar velocity with respect to the ICM \citep{coziol09,lopes18}. Small velocities and high densities are consistent with our observation of moderate bending.

The excess bending of BCGs is approximately the same as for sources between 0.6 and $1.1~r_{500}$, but the central density is much higher. We estimate the density ratio assuming the same generalized NFW profile as in Section~\ref{sec:results_pressure}. This profile diverges at $r=0$, so we instead calculate the ICM density around the BCGs at $r=0.01~r_{500}$ \citep[as in][]{andrade-santos17}. By this estimation, the central density is between 45 and 250 times the density at 0.6 and $1.1~r_{500}$, respectively. If all galaxies in the cluster had motion with the same velocity, then BCGs should be bent the most by ram pressure. Because they are as bent as galaxies where the pressure is 45 to 250 times lower, the BCG velocity offset must be $\sqrt{\rho}=7$ to 16 times lower than the velocity for those galaxies near $r_{500}$. This leads to a typical BCG velocity between 140 and $340~\textrm{km~s}^{-1}$; this is consistent with \cite{lopes18}, who report that 42\% of disturbed clusters in their sample had a BCG with a velocity offset of at least $200~\textrm{km~s}^{-1}$.

We also find that BCGs are more likely to be extended radio galaxies than the average non-BCG, by at least a factor of 2.5. For our sample, the radio luminosity distribution of both BCGs and non-BCGs has a minimum of approximately $2\times 10^{23}~\textrm{W~Hz}^{-1}$. This luminosity is near the upper limit of the radio luminosity function for spiral galaxies \citep{mauch07}, which is evidence that  both sets of galaxies are drawn from a comparable population of ellipticals. This result is consistent with \cite{best07} and \cite{croft07}, who each performed a more detailed analysis of the connection between radio-loud AGN and BCGs, and found that the excess of radio-loud BCGs was dependent on the stellar mass of the galaxy. \cite{best07} reports the excess ranges between two and ten times, depending on the mass bin.

\subsection{Bent sources as tracers of environment}

Because most wide-angle tail (WAT) radio sources are associated with BCGs \citep{odonoghue93}, they can be used to detect clusters \citep[e.g.][]{blanton00}. Various surveys have attempted to use more general populations of bent radio galaxies to detect clusters as well \citep[e.g.][]{obrien16,paterno-mahler17}. However, we find that bent radio sources are not necessarily associated with the \emph{inner} or virialized regions of clusters. At every bending angle, there are more sources beyond $1.5~r_{500}$ than near the cluster center. Observing a strongly bent source is more likely to indicate that it is on the long tail of the distribution of non-cluster galaxy bending angles, as opposed to being due to cluster proximity. Of the 660 highly bent sources ($\Delta\theta>21\degr$) in our sample, 223 (34\%) of them are within the approximately virial region ($1.5~r_{500}$), 455 (69\%) are found out to $10~r_{500}$, and 67 (10\%) are not matched to a cluster within 15~Mpc at all.

We do find that radio sources in general are statistically associated with local overdensities. Within 350~kpc of the radio source, highly bent sources on the outskirts of clusters are located within statistically overdense regions of 4.7 times the background volume density of galaxies; less bent sources are in statistically overdense regions of 2.9 times the background. Other studies of the local companion density around radio-loud galaxies have also detected this excess; for example, \cite{worpel13} found that luminous radio galaxies have almost twice as many optical companions within 160~kpc as did galaxies in their control sample. Now we find that highly bent radio sources in particular have an excess number of companions compared to less bent radio sources, which suggests that even these small overdensities (compared to the densities near the center of clusters) in the local environment are sufficient to induce noticeable bending in the radio jets, as in the model of \cite{venkatesan94}.

In \cite{blanton01}, of 40 visually-identified bent double radio galaxies, 54\% were associated with galaxy clusters and 46\% were associated with galaxy groups. Their sample was more robust than the sample in this work because all the sources had visual confirmation of bent morphology, but their general conclusion is the same as ours: radio galaxies tend to be found in overdense regions, but not necessarily in clusters.

\subsection{Orientation} \label{sec:orientation}

We have used the largest sample of bent radio galaxies compiled so far to probe the orbital distribution of galaxies in the cluster, by measuring the orientation of the radio galaxy tails relative to the cluster. Under the model in which the jets are bent by ram pressure due to relative motion with respect to the ICM, the angle bisector of the opening angle would point away from the direction of motion. \cite{odea85} have previously attempted this same analysis with a study of 70 narrow-angle tail radio sources, and observed a radial preference within a projected 0.5~Mpc of the cluster center but isotropic orbits beyond that.

The distribution of orbital directions in a cluster evolves over time. When a cluster forms via mergers, galaxies and other material tend to fall in with radial orbits; over time, the cluster virializes and the orbits become isotropically distributed. Measuring the degree of isotropy reveals the evolutionary status of the cluster \citep{iannuzzi12}.

The velocity anisotropy parameter $\beta$ is used to parameterize the degree of isotropy by comparing the velocity dispersions in the radial and tangential directions. We use the definition given in \cite{binney87},
\begin{equation} \label{eq:beta}
\beta(r) = 1 - \frac{\sigma_t^2(r)}{\sigma_r^2(r)}.
\end{equation}
The $\beta$ parameter is 0 for isotropically distributed orbits, positive for preferentially radial orbits, and negative for preferentially tangential orbits. We calculated $\beta$ by decomposing the direction of motion of our sample of highly bent sources between 0.01 and $10~r_{500}$ into radial and tangential components, disregarding the position-dependence in Equation~\ref{eq:beta}. We calculate $\beta=0.20\pm 0.02$, consistent with our qualitative assessment of a radial preference.

In our analysis, we stacked our sources by folding the orientation angles around $90\degr$. By doing so we were able to observe the radial preference, but removed any information about if the sources were pointing towards or away from the cluster center. If a pressure gradient (e.g. buoyancy) was the primary driver for bending rather than ram pressure, sources would preferentially open away from the center of the cluster. By unfolding the orientation angles, we observe that sources are preferentially oriented non-tangentially, but do not prefer pointing out rather than in. Therefore, pressure gradients are not the primary driver of bending in our sample.

This is also expected from theoretical considerations. Even in the absence of relative motion, a pressure gradient across the jet will cause it to bend. To first order, the transverse ram pressure force per unit volume in the jet goes as
\begin{equation}
F_r \sim \frac{\rho_{icm} v_{gal}^2}{h},
\end{equation}
as in Equation~\ref{eq:force_balance}. Assuming for simplicity that the ambient pressure gradient is orthogonal to the jet, the force due to it goes as
\begin{equation}
F_p \sim \frac{P_{icm}}{z},
\end{equation}
where $z$ is the pressure scale height. The ratio of the two is simply then
\begin{equation}
\frac{F_r}{F_p} \sim \frac{\rho_{icm} v_{gal}^2}{P_{icm}} \frac{z}{h} \sim M_A^2 \frac{z}{h},
\end{equation}
where $M_A$ is the Mach number of the galaxy's motion relative to the ICM. Statistically, $M_A$ is of order unity, so the main determining factor is the scale height of the pressure gradient. In a relaxed ICM, this will be large, but in a very dynamic ICM, local pressure gradients can be fairly steep ($z$ of order 10~kpc). Therefore, pressure gradients are generally not expected to be the dominant cause of bending, although situations can exist where they are important.

\subsection{Asymmetry}

Whereas \cite{rodman19} used RGZ sources to investigate the role of environment on the length of radio lobes by counting the number of nearby galaxies on both sides of the radio source, we compared the positions of each end of the source (tail or lobe) and the ICM gas pressure at each end using the analytical model of \cite{arnaud10}. We come to the same conclusion as \cite{rodman19} and \cite{malarecki15}: jet expansion into denser environments is suppressed. However, our results are more significant when asymmetry is compared to position in the cluster rather than to the pressure gradient across the source, suggesting that either the density gradient is the more important contribution or that an important piece is being left out of our simplified pressure model. In addition, \cite{rodman19} specifically looked at FR-II sources, whereas our population contains both FR-I and FR-II sources. Tail length is not well defined for FR-I sources, and there may be physical differences between the propagation into the ambient medium of the diffuse lobes of FR-I sources compared to the more collimated tails of FR-II sources, so the inclusion of FR-I sources in our sample may have reduced the signal.

\section{Conclusion}

In this work we investigated the effect of cluster environment on the morphology of a sample of 4304 radio galaxies from Radio Galaxy Zoo. We found that clusters do exert a statistical influence on the bending of radio tails; namely, radio galaxies tend to be more bent closer to the center of the cluster, where the ICM density is higher. They are also more bent in higher mass clusters, where the relative velocity between the galaxy and the ICM is higher. Together, this leads to the conclusion that we are observing the influence of ram pressure exerted on the radio galaxy by the ICM, which is further supported by connecting the ram pressure to the ICM gas pressure, which also correlates with bending. We also look at the use of radio galaxies as a tracer for cluster proximity; more than half of highly bent radio galaxies in our sample are farther from clusters than $1.5~r_{500}$, the limit of the cluster's statistical influence on radio bending. On the other hand, we look at a sample of highly bent radio galaxies far from the nearest cluster, and find that these galaxies are still associated with local local overdensities in the population of field galaxies.

We were unable to investigate the relationship between the physical extent of the radio emission and cluster environment, as any correlation between physical size and bending was incidentally removed as part of the bending error correction for angular size in Section~\ref{sec:correction}. Even prior to the correction, redshift biases confounded our tentative claim that sources are physically smaller inside clusters. As we showed with our radio tail asymmetry analysis, jet propagation is somewhat suppressed in a denser environment, so we hypothesize that the radio sources will be physically smaller. Thus, smaller sources would also be more bent, which is unfortunately degenerate with the sources of error in our bending measurement. Future work with a larger sample of radio galaxies is necessary to be able to tease apart this association from the measurement uncertainty. As machine learning algorithms are developed to identify and classify radio galaxies from forthcoming large-scale radio surveys such as EMU \citep{norris11} and the VLASS \citep{myers18}, and much more extensive optical surveys such as Pan-STARRS \citep{chambers16} become available for host searches, we should be able to identify second-order correlations such as this. Machine learning may also lead to a more physically significant definition of what qualities define bending angle, rather than the approximation used in this work.

We plan to follow up this paper with a detailed look at the relaxation states of the clusters, and how that impacts our bending results. In particular, we will be able to investigate whether disturbed clusters contain BCGs that have more distorted radio morphologies, and whether a connection between the distribution of velocity vector orientations derived from the radio bending and the relaxation state can be measured.

\acknowledgments

This publication has been made possible by the participation of more than 12,000 volunteers in the Radio Galaxy Zoo project. Their contributions are individually acknowledged at \url{rgzauthors.galaxyzoo.org}. We thank the anonymous referee for their constructive feedback in improving the manuscript.

Partial support for this work for A.F.G., L.R., and T.W.J. comes from National Science Foundation grant AST-171405 to the University of Minnesota. The support for J.-A.K. comes from the National Research Foundation of Korea to the Center for Galaxy Evolution Research through the grant program No. 2017R1A5A1070354. H.A. benefited from University of Guanajuato grant DAIP \#66/2018. F.dG. is supported by the VENI research program with project number 639.041.542, which is financed by the Netherlands Organisation for Scientific Research (NWO). S.S.S. thanks the Australian Research Council for an Early Career Fellowship, DE130101399.

This publication makes use of radio data from the Karl G. Jansky Very Large Array, operated by the National Radio Astronomy Observatory. The NRAO is a facility of the National Science Foundation operated under cooperative agreement by Associated Universities, Inc. This publication makes use of data products from the \emph{Wide-field Infrared Survey Explorer}, which is a joint project of the University of California, Los Angeles, and the Jet Propulsion Laboratory/California Institute of Technology, funded by the National Aeronautics and Space Administration. Funding for the Sloan Digital Sky Survey IV has been provided by the Alfred P. Sloan Foundation, the U.S. Department of Energy Office of Science, and the Participating Institutions. SDSS-IV acknowledges support and resources from the Center for High-Performance Computing at the University of Utah. The SDSS web site is \url{www.sdss.org}. This research made use of \texttt{Astropy}, a community-developed core Python package for Astronomy \citep{astropy13, astropy18}, and packages in the \texttt{SciPy} ecosystem \citep{jones01}.

\appendix

We present two supplementary data tables: one consisting of the 3723 sources matched to WH15 within a projected separation of 15~Mpc and redshift difference of $\pm 0.04(1+z)$, and the other consisting of the 581 sources \emph{not} matched to WH15. The full, machine-readable version of these tables are available; a portion of the matched table is shown in Table~\ref{tab:supplement} for guidance on form and content. The unmatched table is formatted the same, but without columns 14 through 24. A description of all parameters and columns is as follows:

\emph{Column 1:} Radio Galaxy Zoo source name.

\emph{Column 2:} Zooniverse subject name. The radio/IR overlays used for classification can be found at \url{radiotalk.galaxyzoo.org/#/subjects/ARG000****}, where \texttt{ARG000****} is the Zooniverse subject name.

\emph{Column 3:} Radio morphology. Double sources are denoted by 2 and triple sources are denoted by 3.

\emph{Column 4:} Largest angular size ($LAS$) of the source, defined as the diagonal of the bounding box that contains the outermost ($4\sigma$) FIRST contours, in units of arcminutes. The error in source size is dominated by the arbitrary contour threshold in the FIRST images and would require detailed modeling to estimate.

\emph{Column 5:} R.A. (J2000) of the SDSS host galaxy.

\emph{Column 6:} Decl. (J2000) of the SDSS host galaxy.

\emph{Column 7:} Redshift of the SDSS host galaxy.

\emph{Column 8:} Uncertainty in the redshift of the SDSS host galaxy.

\emph{Column 9:} Type of redshift used. Spectroscopic redshifts were used when available and are denoted by $s$; otherwise, photometric redshifts were used and are denoted by $p$.

\emph{Column 10:} Measured bending angle ($\theta$) of the source, in units of degrees. The error in $\theta$ and derived quantities ($\theta_{corr}$, $\Delta\theta$, and $\phi$) can only be determined statistically, as discussed in Section~\ref{sec:uncertainty}.

\emph{Column 11:} Corrected bending angle of the source ($\theta_{corr}$), in units of degrees. See Section~\ref{sec:correction} for more information.

\emph{Column 12:} Excess bending angle of the source ($\Delta\theta$), in units of degrees. See Section~\ref{sec:correction} for more information.

\emph{Column 13:} Source asymmetry (Asym.), defined as the ratio of the lengths of the inward-pointing tail to the outward-pointing tail. For sources not matched to clusters, a tail was randomly selected as the ``inward-pointing'' one. See Section~\ref{sec:results_asym} for more information.

\emph{Column 14:} WH15 cluster name.

\emph{Column 15:} R.A. (J2000) of the WH15 cluster.

\emph{Column 16:} Decl. (J2000) of the WH15 cluster.

\emph{Column 17:} Redshift of the WH15 cluster.

\emph{Column 18:} Type of redshift used. Spectroscopic redshifts were used when available and are denoted by $s$; otherwise, photometric redshifts were used and are denoted by $p$.

\emph{Column 19:} Separation ($r$) from the nearest WH15 cluster, in units of $r_{500}$. Because separations less than $0.01~r_{500}$ are not physically meaningful (see Section~\ref{sec:sample_dist}), we have truncated the values to two decimal points. The error in $r$, assuming a correct host identification, is limited by the positional and redshift uncertainty in SDSS.

\emph{Column 20:} Radius ($r_{500}$) of the nearest WH15 cluster, in units of Mpc.

\emph{Column 21:} Mass ($M_{500}$) of the nearest WH15 cluster, in units of $10^{14}~M_\sun$.

\emph{Column 22:} Logarithm of the local ICM gas pressure ($P_{icm}$), in units of $\log_{10}~\textrm{keV~cm}^{-3}$. See Section~\ref{sec:results_pressure} for more information.

\emph{Column 23:} Orientation ($\phi$) of the source opening angle with respect to the nearest WH15 cluster, in units of degrees. This is the ``unfolded'' orientation angle discussed in Section~\ref{sec:results_orient}, where sources with orientation $\phi<90\degr$ open away from the cluster and sources with orientation $\phi>90\degr$ open towards the cluster.

\emph{Column 24:} Alignment (Align.) of the source tails with respect to the cluster. Radially-aligned sources are denoted by $r$, tangentially-aligned sources are denoted by $t$, and all other sources are denoted by $o$. See Section~\ref{sec:results_asym} for more information.




\bibliographystyle{aasjournal}
\bibliography{main} 



\begin{longrotatetable}
\begin{splitdeluxetable*}{cccDDDDDcDDDDBcDDDcDDDDDc}
\tablenum{2}
\tablecaption{The first 10 rows of the cluster-matched data set used in this work.}
\label{tab:supplement}
\tablehead{ \colhead{RGZ name} & \colhead{Zooniverse ID} & \colhead{Morph.} & \twocolhead{$LAS$} & \twocolhead{R.A.} & \twocolhead{Decl.} & \twocolhead{$z$} & \twocolhead{$z_{err}$} & \colhead{$z_{type}$} & \twocolhead{$\theta$} & \twocolhead{$\theta_{corr}$} & \twocolhead{$\Delta\theta$} & \twocolhead{Asym.} & \colhead{WH15 name} & \twocolhead{R.A.} & \twocolhead{Decl.} & \twocolhead{$z$} & \colhead{$z_{type}$} & \twocolhead{$r$} & \twocolhead{$r_{500}$} & \twocolhead{$M_{500}$} & \twocolhead{$P_{icm}$} & \twocolhead{$\phi$} & \colhead{Align.} }
\decimalcolnumbers
\startdata
J000024.0+122953 & ARG0002uly & 2 & 0.764 & 0.10026 & 12.49813 & 0.6559 & 0.0966 & $p$ & 6.9 & 6.8 & -2.6 & 1.40 & J000055.3+123153 & 0.23058 & 12.53137 & 0.5936 & $s$ & 4.99 & 0.65 & 9.28 & -5.69 & 37.6 & $t$ \\
J000042.3+143545 & ARG0002p8y & 2 & 0.242 & 0.17650 & 14.59587 & 0.6223 & 0.0002 & $s$ & 79.5 & 39.4 & 38.8 & 1.25 & J000042.4+143545 & 0.17650 & 14.59587 & 0.6224 & $s$ & 0.00 & 0.71 & 10.49 & 0.16 & 59.8 & $o$ \\
J000054.8+032536 & ARG0003lyd & 2 & 0.445 & 0.22854 & 3.42676 & 0.5356 & 0.0002 & $s$ & 34.6 & 22.5 & 21.2 & 1.19 & J000054.8+032536 & 0.22854 & 3.42676 & 0.5334 & $s$ & 0.00 & 0.68 & 9.66 & 0.13 & 3.3 & $t$ \\
J000202.5+143333 & ARG0002pc8 & 2 & 0.756 & 0.51050 & 14.55944 & 0.5629 & 0.0327 & $p$ & 14.5 & 14.3 & 12.3 & 0.87 & J000136.3+142125 & 0.40115 & 14.35699 & 0.5859 & $p$ & 8.23 & 0.68 & 10.13 & -6.75 & 69.0 & $r$ \\
J000333.4+033647 & ARG0003lfy & 3 & 1.025 & 0.88945 & 3.61321 & 0.0849 & 0.0080 & $p$ & 2.5 & 3.5 & -6.4 & 0.81 & J000318.1+043739 & 0.82531 & 4.62755 & 0.0963 & $s$ & 7.17 & 0.94 & 14.71 & -6.66 & 35.3 & $t$ \\
J000551.6+120312 & ARG0002w4j & 2 & 0.294 & 1.46528 & 12.05344 & 0.3118 & 0.1153 & $p$ & 1.1 & 0.6 & -7.3 & 0.95 & J000524.4+115330 & 1.35161 & 11.89157 & 0.3154 & $s$ & 6.34 & 0.53 & 3.27 & -6.69 & 90.5 & $r$ \\
J000608.3+121431 & ARG0002vh4 & 2 & 0.545 & 1.53466 & 12.24224 & 0.4983 & 0.0507 & $p$ & 3.2 & 2.4 & -6.9 & 1.52 & J000610.5+121358 & 1.54388 & 12.23287 & 0.4969 & $p$ & 0.52 & 0.60 & 7.48 & -2.15 & 82.1 & $r$ \\
J000651.3+070352 & ARG0003bdt & 2 & 0.571 & 1.71414 & 7.06440 & 0.6185 & 0.0848 & $p$ & 1.7 & 1.3 & -7.2 & 0.60 & J000511.8+065821 & 1.29916 & 6.97248 & 0.6632 & $s$ & 16.55 & 0.66 & 9.72 & -8.27 & 104.8 & $r$ \\
J000728.1+110618 & ARG0002z9e & 2 & 0.975 & 1.86720 & 11.10506 & 0.6729 & 0.0663 & $p$ & 3.6 & 4.8 & -5.5 & 0.83 & J000940.0+111627 & 2.41657 & 11.27403 & 0.6768 & $s$ & 13.65 & 1.08 & 34.19 & -7.46 & 114.4 & $r$ \\
J000819.6+115449 & ARG0002wkx & 2 & 0.569 & 2.08194 & 11.91383 & 0.1744 & 0.0210 & $p$ & 52.3 & 40.1 & 39.5 & 0.74 & J001010.1+121811 & 2.54219 & 12.30297 & 0.1745 & $s$ & 7.11 & 0.92 & 15.56 & -6.58 & 53.4 & $o$ \\
\enddata
\tablecomments{The full, machine-readable version of this table is available on the journal website. A portion is shown here for guidance on form and content.}
\end{splitdeluxetable*}
\end{longrotatetable}

\end{document}